\documentclass[twocolumn,showpacs,prb,amsmath,amssymb]{revtex4}

\usepackage{graphicx}\usepackage{dcolumn}\usepackage{bm}

\begin{document}

\title{Luttinger liquids with curvature:
Density correlations and Coulomb drag effect}
\author{D. N. Aristov}
\altaffiliation[On leave from ]
{Petersburg Nuclear Physics Institute, Gatchina  188300, Russia.}
\affiliation{
Institut f\"ur Theorie der Kondensierten Materie, Universit\"at
Karlsruhe, 76128 Karlsruhe, Germany
}
\affiliation{
Center for Functional Nanostructures, Universit\"at Karlsruhe,
76128 Karlsruhe, Germany}
\date{\today}

\begin{abstract}
We consider the effect of the curvature in fermionic dispersion on the
observable properties of Luttinger liquid (LL).
We use the bosonization technique where
the curvature is irrelevant perturbation, describing
the decay of LL bosons (plasmon modes).
When possible, we establish the correspondence between the
 bosonization and the fermionic approach.
We analyze modifications in density correlation functions
due to curvature at finite temperatures, $T$.
The most important application of our approach is the analysis of the
Coulomb drag by small momentum transfer between two LL, which is only
possible due to curvature.
Analyzing the a.c. transconductivity in the one-dimensional drag setup,
we confirm the results by Pustilnik {\em et al.} for $T-$dependence
of drag resistivity,  $R_{12} \sim T^2$ at high and $R_{12} \sim T^5$ at
low temperatures.
The bosonization allows for treating both intra- and
inter-wire electron-electron interactions in all orders, and we
calculate exact prefactors in low-$T$ drag regime.
The crossover temperature between the two regimes is $T_1 \sim E_F
\Delta$, with $\Delta$ relative difference in plasmon velocities. We
show that $\Delta \neq 0$ even for identical wires, due to lifting
of degeneracy by interwire interaction, $U_{12}$, leading to crossover
from $R_{12} \sim U_{12}^2 T^2$ to $R_{12} \sim T^5/U_{12}$
at $T \sim U_{12}$.
\end{abstract}

 \maketitle

\section{Introduction}

The effects of the curvature of fermionic dispersion for the
observables in strongly-correlated one-dimensional (1D) fermionic
systems were discussed by several groups recently.
\cite{PMGA,AbaWieg05,PKKG06,Teber06,G7,Kopietz05,Gutman06}
One physical effect, which probes both the curvature and
interactions is the Coulomb drag effect.
In a typical experimental setup, this effect is observed as non-zero
drag resistivity, $R_{12} = V_2/I_1$, with a d.c.
current $I_1$ flowing in the "active" wire and the voltage
bias $V_2$ per unit length applied to the second "passive" wire in
order to assure $I_2 =0$, see review articles Ref.\
[\onlinecite{reviews}].

It is generally agreed that the principal source of the drag
effect is the particle-hole asymmetry of the electronic system.
This was established in various works considering electrons in higher
spatial dimensions, $D>1$, both in the absence of magnetic field
\cite{drag90s} and in the presence of it (see \cite{GoMivOp04} and
references therein).
It was shown that the leading contribution to the drag effect is
obtained in the second order of the interwire interaction $U_{12}$
and is schematically depicted by the Feynman graph in Fig.
\ref{fig:drag1st}. This
contribution corresponds to the virtual processes away from the Fermi
surface, and is ultimately determined by the curvature of the
electronic dispersion.

One naturally would expect that the curvature should be also
responsible for the drag effect in one spatial dimension (1D).
However, the first study of the drag effect in 1D was devoted to
a different mechanism, \cite{NazAve98} specific for
1D and requiring some additional conditions, we return to it below.
The study of the drag effect in 1D due to curvature of electronic
dispersion was initiated only recently, by Pustilnik et al.\ \cite{PMGA}
The reason is that the drag effect depends both on the curvature and the
fermionic interaction, whereas it is known that the
interaction is very important in 1D and ultimately leads to the notion
of Luttinger liquid. The possibility to find a complete solution, i.e.
Luttinger liquid, for the interacting system, lies in the crucial
simplification of the theory - the linearization of the fermionic
dispersion around the Fermi energy. In most cases it is an innocent
procedure, and does not influence the final result. The Coulomb drag
effect is one example, where this usual theoretical trick with
linearization leads to the immediate disappearance of
the observable quantity in question. In their work, \cite{PMGA}
Pustilnik et al.\ (see also \cite{PKKG06}) hence used the fermionic
formalism, and found the drag effect as the function of temperature, for
different regimes. Restricting themselves to the lowest order of
interwire, as well as intrawire
interactions, they showed that $R_{12} \propto T^2$ for equal wires and
$R_{12} \propto T^5$ for non-equal wires. In this work we
check their results by another method and extend their treatment beyond
the lowest order of perturbation theory (PT) in fermionic interactions.

To this end, we use the traditional tool for studying the
effects of interaction in 1D, known as bosonization technique,
which is devised for effective resummation of the appearing series
in PT. \cite{GoNeTs}
Employing the bosonization we take into account the inter- and
intrawire interaction in all orders, whereas the curvature of
fermionic dispersion is treated as perturbation.
We show that it suffices to consider fourth order in curvature when
calculating low-temperature drag effect. At higher temperatures we
perform the resummation of most singular contributions in PT
in curvature, incorporating free-fermion results into bosonization
language.
In this sense, our method is complementary to the analysis in
\cite{PMGA} where the curvature was treated exactly and the
interactions --- in the lowest necessary order.

The issue of the curvature of fermionic dispersion in bosonization
technique was rarely discussed previously. Therefore, although
we are ultimately interested in the drag effect, we develop a
systematic approach to Luttinger liquid with curvature.
It was shown long ago
\cite{Schick68,Haldane81} that the curvature
corresponds to the cubic terms in bosonic densities (plasmon modes)
of so-called right- and left-moving fermions near
two Fermi points $\pm k_F$, respectively. The appearance
of these cubic terms corresponds to the decay of bosons (plasmons), and
was previously derived in two ways via operator identities. We suggest
yet another explanation for the appearance of the cubic
terms in the theory, an explanation referring to Green's function
formalism and more pertinent to our discussion of Coulomb drag.
Besides that we discuss the modification of the current
operator, dynamic density correlations in isolated wires, and the
light-cone singularities at finite temperatures $T$. Keeping $T$ finite
is important for our calculation, as the drag effect is only due to
inelastic processes of fermionic interaction, and hence vanishes at
$T=0$. We use the Kubo formalism, which technically delivers
the optical transconductivity, $\sigma_{ij}(\omega)$, in one dimensional
system of two wires, and eventually extract the d.c. drag
resistivity, $\sigma_{ij}^{-1}(0)$. Given our assumption of clean
Luttinger liquids, it is a delicate procedure and we discuss it in
detail.

The plan for the paper is as follows. We set up the problem and discuss
the identification of fermionic curvature with decay of plasmon modes
in Sec.\ \ref{sec:setup}. The form of density correlation functions is
discussed in Sec.\ \ref{sec:density}, we show the correspondence
between fermionic approach and bosonization here. The
bosonization expression for current and conductivity for one wire is
discussed in Sec.\ \ref{sec:conduc1}. The transconductivity matrix for
two wires is considered in Sec.\ \ref{sec:conduc2}, and the drag
coefficient is obtained for relatively simple case of non-equal wires.
The analysis of nearly-equal wires is performed separately, in most
technically involved Sec.\ \ref{sec:nearlyequal}, the final results
for $R_{12}$ in low temperature regime are discussed here. We present
our summary and conclusions in Sec.\ \ref{sec:discussion}.

\section{setting up the problem}
\label{sec:setup}
\subsection{Hamiltonian}

We consider spinless (spin-polarized) electrons in two wires, with the
forward-scattering short-range interaction both inside and between
the wires. The electrons in $i$th wire ($i=1,2$) are conventionally
subdivided into right- and left-going species, centered around
corresponding Fermi momenta, $k_{Fi} = \pi n_i$, with the
fermionic density $n_i$. We go beyond the the linearized spectrum
approximation, keeping quadratic terms in dispersion.

As usual, we decompose fermionic operator into "slowly oscillating"
chiral components
            \begin{equation}
            \psi(x) = e^{ik_Fx} \psi_R(x) + e^{-ik_Fx} \psi _L(x),
            \end{equation}
so that the smooth part of the fluctuating fermionic density is
            \begin{equation}
            \rho(x) = \psi^\dagger _R(x)\psi_R(x) +
            \psi^\dagger _L(x)\psi_L(x).
            \end{equation}

Our Hamiltonian is given by four terms :
            \[
            {\cal H}  = \int dx\,
            (H_1 + H_2 + H_{12} + H_{cur}),
            \]
with the linearized interacting fermions in individual wires
            \begin{eqnarray}
            H_{j=1,2} &=&
            -i v_{Fj}\psi^\dagger _{Rj}
            \partial_x\psi_{Rj}
            + i v_{Fj} \psi^\dagger _{Lj} \partial_x\psi _{Lj}
            \nonumber \\ &&
            + U_j \rho_j^2(x) /2,
            \label{ferHam}
            \end{eqnarray}
forward-scattering interaction between wires
            \begin{equation}
            H_{12} =  U_{12}
            \rho_1(x) \rho_2(x),
            \end{equation}
and the terms describing the curvature of fermionic dispersion
            \begin{equation}
            H_{cur} =  \sum_{j=1,2}
            \frac{1}{2m_j}
            \left( \psi^\dagger _{Rj} (i\partial_x)^2 \psi_{Rj}
            +\psi^\dagger_{Lj} (i\partial_x)^2\psi _{Lj}
            \right).
            \end{equation}

The bosonization approach to the above system is the
application of the important representation of 1D chiral fermions as
            \begin{equation}
            \psi_{Rj} = \frac1{\sqrt{2\pi\Lambda}} e^{i\phi_{Rj}} ,
             \quad
            \psi _{Lj} = \frac1{\sqrt{2\pi\Lambda}} e^{-i\phi_{Lj}} ,
            \label{fermion}
            \end{equation}
with $\Lambda$ the ultraviolet cutoff. Here he
bosonic fields $ \phi_{R(L),j}= \phi_j \mp \theta_j$, with primary field
$\phi_j$ and its canonically conjugated momentum $\Pi_j =
\pi^{-1}\partial _x \theta_j$ satisfying $[\phi_j(x),\Pi_l(y)] = i
\delta_{jl} \delta(x-y)$.

In this bosonization notation \cite{GoNeTs}, the Hamiltonian of the
system is rewritten  as follows :
    \begin{eqnarray}
     {\cal H}   &=& \int dx\,( H_1 + H_2 + H_{12} + H_{cur} ) ,
     \label{ham} \\
     H_{i=1,2} &=&
     \frac{1}{2\pi}\left(
     v_{iJ} (\partial_x \theta_i)^2 + v_{iN} (\partial_x
 \phi_i)^2 \right)
     , \\
    H_{12}&=&
    \frac{U_{12}}{\pi^2}  \partial_x \phi_1 \partial_x \phi_2,
 \\
     H_{cur} &=&
     \frac 1 {6\pi m_1}
     \partial_x \phi_1 \left(3 (\partial_x \theta_1)^2 + (\partial_x
 \phi_1)^2
     \right ) + (1\leftrightarrow 2) .
     \nonumber
    \end{eqnarray}
Here and below I use the shorthand notation $\partial_x \phi_1 =
 \partial
 \phi_1(x) /\partial x$, etc.
The electronic density operator in (\ref{ham}) is given by $\rho_i =
 \partial_x  \phi_i/\pi$.
We follow the notation by Haldane, with $v_{iJ} =
 v_{Fi} = k_{Fi} /m_i$, $v_{iN}  = v_{Fi} + U_i/\pi$. In the absence of
$H_{12}$, $H_{cur}$,  the dispersion of  (plasmon) excitations is
$\varepsilon_{i}(q)
= v_i |q|$, where $v_i^2 = v_{iJ} v_{iN}$. The strength of the intrawire
interactions may  be  encoded in the Luttinger parameters, $K_i =
\sqrt{v_{iJ}/ v_{iN}}$ ; it is convenient also to define
the dimensionless interwire interaction, $u = \pi^{-1}U_{12}/
 \sqrt{v_{1N}v_{2N}}$.

It is seen that the Hamiltonian (\ref{ham}) contains an exactly
solvable quadratic-in-bosons part $ H_1 + H_2 + H_{12}$ and the
curvature acts as the interaction.

\subsection{Curvature as interaction}

In bosonization, the normally ordered parts
of chiral fermionic densities become
            \begin{equation}
            R_j \equiv \psi^\dagger _{Rj}\psi_{Rj}  ,\quad
            L_j \equiv \psi^\dagger _{Lj}\psi_{Lj},
            \end{equation}
with  $\rho_j  = R_j + L_j = \pi^{-1} \partial_x \phi_j$.
The main statement of (Abelian) bosonization, dating back to Tomonaga,
\cite{Tomonaga} is that the free chiral fermions with linear dispersion
are equivalent to quadratic form in chiral densities :
$ -i \psi^\dagger _{R} \partial_x\psi_{R} = \pi R ^2$.
Tomonaga established this relation, studying the equation of motion for
the chiral density $R$ with the fermionic Hamiltonian with
linear dispersion. This observation was later corroborated by Schick,
\cite{Schick68} who found that similar consideration with the quadratic
fermionic dispersion yields another important relation :
      \begin{equation}
      \psi^\dagger _{R}
      \frac{(i\partial_x)^2}{2m}
      \psi_{R}
      = \frac{2\pi^2}{3m} R^3.
      \label{cubic}
      \end{equation}

Another derivation of (\ref{cubic}) was shown in the paper by Haldane,
\cite{Haldane81}  who started with the fermion representation
(\ref{fermion}). Note that using
Eq.\  (\ref{fermion}), one arrives also to the term in the Hamiltonian,
which constitutes a full derivative, $\sim \partial_x^2 \phi_R
\partial_x \phi_R \sim \partial_x (R^2)$, which does not
contribute to the dynamics. Confining oneself with the quadratic terms
in the fermionic dispersion, one obtains nothing beyond the terms $R^2 +
L^2$ and $R^3 + L^3$ in the bosonization language. The stability
issue for the cubic action is briefly discussed in the next section.

\begin{figure}
\includegraphics[width=8cm]{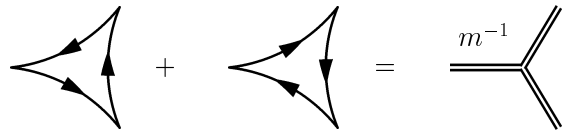}%
\caption{Feynman diagrams representing non-linear response of
fermionic system. Two three-tail fermion loops correspond
to a boson decay process, see the text for discussion.
\label{fig:threetail}}
\end{figure}

\begin{figure}
\includegraphics[width=7cm]{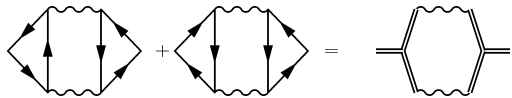}%
\caption{First Coulomb drag diagrams in the fermionic formalism and
their bosonization counterpart, see the text for discussion.
\label{fig:drag1st}}
\end{figure}

Keeping in mind a problem of drag effect, let us provide yet
another
argument verifying the form of the curvature term in bosonization. It is
well known that the multi-tail fermionic loop diagrams in one spatial
dimension are exactly zero for the linear dispersion law. This
statement is known as the higher-loop cancellation theorem after the
work by Dzyaloshinskii and Larkin. \cite{DzyaLar73} One can show
however, that the
presence of the curvature in fermionic dispersion leads to generally
non-zero value for the processes given by triangle diagram.  The
expression for this diagram, characterized by three external vertices,
can be easily calculated at $T=0$ as
\begin{eqnarray}
    {\cal T}&=&
    \frac{q_1 q_2 q_3
    \left(\sum_{i=1}^3\Lambda(\omega_i,q_i)\right)}
    {(\omega_1 q_2 -\omega_2 q_1 )^2 - (q_1q_2q_3/2m)^2},
    \label{triangle-full}
    \\
    \Lambda(\omega,q)&=&\frac{1}{2\pi}
    \ln\frac{\omega - v_Fq+ q^2/2m }{\omega - v_Fq- q^2/2m},
\end{eqnarray}
with $q_3 = q_1+q_2$, $\omega_3 = \omega_1+  \omega_2$. One can check
that ${\cal T}$ vanishes in the limit $m\to \infty$.
Expanding
(\ref{triangle-full})
to the leading order of $m^{-1}$, one obtains the structure
 \begin{equation}
  \frac1{2\pi m} \frac{q_1}{v_Fq_1 - \omega_1}
 \frac{q_2}{v_Fq_2 -\omega_2}\frac{q_3}{v_Fq_3 -
\omega_3},
  \label{triangle-red}
 \end{equation}
which corresponds to three density
correlators for linear spectrum, $(2\pi)^{-1}q/(v_Fq - \omega)$,
attached to above vertex, $2\pi^2/(3m)$; factor $3! = 6$ comes from
symmetrization. Schematically, it is shown as Feyman diagrams in Fig.\
\ref{fig:threetail}, where fermion Green's functions are shown by
lines with arrows, and double lines represent bosonic
Green's functions (\ref{bare-g}).
Note that the leading-order drag diagram is given by the "two
triangles" fermionic diagrams in Fig.\ \ref{fig:drag1st},
with wavy lines standing for $U_{12}$.\cite{drag90s,GoMivOp04} Comparing
it to  Fig.\ \ref{fig:threetail} we see that the drag coefficient
should
correspond to "two stars" bosonic diagram in the r.h.s. of Fig.\
\ref{fig:drag1st}. Further observing that interwire interaction,
$U_{12}$, is exactly included into the Luttinger liquid formalism, Eq.\
(\ref{ham}), we eventually arrive to the diagrams depicted in
Fig.\ \ref{fig:123} below. Note that passing from Fig.\
\ref{fig:drag1st} to Fig.\ \ref{fig:123} corresponds to the screening of
interwire interaction, see e.g.\ Kamenev and Oreg. \cite{drag90s} The
RPA resummation in case of LL gives an exact result; however, the
simplest bosonic diagram in Fig.\ \ref{fig:drag1st}
corresponds to the optical transconductivity, rather than
to d.c.\ drag effect, as will be shown below.

Before considering a general case of two different wires, let us
discuss the modification of density correlations in one wire with the
curvature term.

\section{density correlations}
\label{sec:density}

Let us first discuss the density correlation function for one wire,
with and without the Luttinger-liquid type interaction. Such analysis
was previously done in \cite{Teber06,G7}, we extend it
for the case $T\neq 0$.  The bosonized Hamiltonian reads as
    \begin{eqnarray}
    {\cal H} &=& \int dx\, (  H_0 +{H}_{cur}),
    \label{Ham1wire}
    \\
     {H}_0 &=& \frac{1}{2\pi}\left( v_{J}
          (\partial_x\theta)^2 + v_{N}(\partial_x \phi)^2 \right), \\
      &=&    \pi  v_F (R^2 + L^2) +
         \frac{U}{2}(R+L)^2 ,
         \label{onewireHam} \\
     H_{cur}  &=&
     \frac 1 {6\pi m}
     \partial_x \phi \left(
     3 (\partial_x \theta)^2 + (\partial_x \phi)^2
     \right ) ,  \\
      &=&
     \frac{2\pi^2}{3m}  \left( R^3 + L^3 \right ) .
     \label{cubicterm}
    \end{eqnarray}
The commutation rules for the chiral densities are
    \begin{eqnarray}
     [R (x), R(y)] &=& -   [L (x), L(y)] =
     \frac{i}{2\pi}\partial_x \delta(x-y).
     \label{KacMoody}
    \end{eqnarray}
Our choice of the interaction in the form $U (R+L)^2$
corresponds to the case $g_4 = g_2$ in more general form of
the Hamiltonian $g_4(R^2+L^2) + 2g_2 RL$.

\subsection{Free fermions}

In the absence of the mixing interaction term $U RL$ in
(\ref{onewireHam}), each chiral component is considered separately.
In the absence of the curvature, $1/m =0$, the bare boson retarded right
(left) Green's functions is independent of temperature and given by
      \begin{equation}
      g^{(0)}_{R(L)} = \frac1{2\pi} \frac{q}{qv_F \mp (\omega + i0)},
      \label{bare-g}
      \end{equation}
with $\delta-$function spectral weight.

Importantly, in this free case we know the exact expression for
correlation function of the right (left) moving densities. The retarded
(right) density propagator $\chi(\omega, q)$ can be calculated at
non-zero temperature as the fermionic loop diagram,
      \begin{eqnarray}
     \chi &=&      \int \frac{dk}{4\pi}
     \frac{\tanh(\varepsilon_k/2T) - \tanh(\varepsilon_{k+q}/2T)}
     {\omega + \varepsilon_k - \varepsilon_{k+q}+ i0}
    \end{eqnarray}
with $\varepsilon_k = v_F k + k^2/2m$.
Neglecting the terms of order of $T/(mv_F^2)$, we get ~:
     \begin{eqnarray}
     {\rm Im}\;\chi &=& \frac{m}{4q}
     \frac{\sinh(\omega/2T)}
     {\cosh\left(\frac{qv_F(\Omega-1)}{4T}\right)
      \cosh\left(\frac{qv_F(\Omega+1)}{4T}\right)},
      \\
     \chi &=& \frac{q}{4\pi w_q} \left[
     \psi\left(\frac12 - i \frac{|qv_F|(\Omega-1)}{4\pi T} \right)
     \right. \nonumber \\ && \left.
     -\psi\left(\frac12 - i \frac{|qv_F|(\Omega+1)}{4\pi T} \right)
      \right],
      \label{exacpropT}  \\
      \Omega &=& (\omega- qv_F)/w_q, \quad
      w_q= q^2/(2m),
    \end{eqnarray}
with $\psi(x)$ digamma function. Restoring the real part of $\chi$ we
used the rule
\[ i \tanh z \to -\frac{2}{\pi}
\psi\left(\frac12 - i \frac{z}{\pi} \right) \]
for the function analytic in upper semiplane of $z$.
In the limit of zero temperature we
obtain
     \begin{eqnarray}
     {\rm Im}\;\chi &=& \frac{m}{2q} \Theta( 1- |\Omega|) \\
      \chi &=& \frac{q}{4\pi w_q}
       \ln \frac {\omega- qv_F  -w_q+i0}{ \omega- qv_F +w_q+i0},
     \label{exacpropT=0}
    \end{eqnarray}
with the step-function $\Theta(x) =1$ at $x>0$.

Let us now discuss how these exact expressions relate to the
bosonization theory. Formally, the curvature term $H_{cur}$ has a
scaling dimension 3, which is strongly irrelevant.
The common wisdom then
would prescribe to treat it in the lowest order of perturbation theory
(PT) in the infrared limit, $q,\omega \to0$. Comparing
(\ref{exacpropT=0}) and the bosonization expression (\ref{bare-g}), one
can see that the $\delta-$function is a good approximation for the
spectral weight in this limit. However, the
lowest PT contribution, with two bosons in intermediate state,
leads to the self-energy part of the form
      \begin{eqnarray}
     \Sigma &=& 2\pi \frac{q}{12m^2}
     \frac {q^2+ (2\pi T/v_F)^2}{qv_F - \omega-i0},
     \label{Sigma}
    \end{eqnarray}
Note, that in a quite unusual way \cite{Samo98} the spectral weight
${\rm Im}\;\Sigma(\omega, q)$ is located exactly at the "light cone" of
bare spectrum, $\omega - qv_F = 0$.
The Dyson equation, $g^{-1}_R = (g^{(0)}_R)^{-1} -
\Sigma$, reads as
       \begin{eqnarray}
     g^{-1}_R = 2\pi\left[ \frac{qv_F - \omega }{q} -
      \frac{q}{12m^2}
     \frac {q^2+ (2\pi T/v_F)^2}{qv_F - \omega}
     \right].
     \label{DysonEq}
    \end{eqnarray}
It means that instead of previous single pole $\omega = qv_F$, we
have two poles separated by a distance $\sim m^{-1} q \max[q, T]$.

Now we compare (\ref{DysonEq}) with (\ref{exacpropT}), by expanding
$\chi^{-1}$ in powers of $m^{-1}$. We find that (\ref{DysonEq}) is
reproduced by (\ref{exacpropT}) in the order $m^{-2}$. It suggests
that bosonization produces a correct {\em asymptotic} series in
$m^{-1}$, but at the energies close to "light-cone" condition, an ever
increasing number of PT corrections should be taken into account, to
arrive at the exact answer (\ref{exacpropT}).

Interestingly, the bosonization predicts also that the
fermionic interaction of the form $g_4 (R^2 + L^2)$, i.e.\ with $
g_2=0$, results only in the
shift of the Fermi velocity, and therefore the spectral
weight $\chi$ has the previous form (\ref{exacpropT})
except for the change, $v_F\to v_F +g_4/2\pi$. This statement
should be contrasted with a recent work
\cite{Kopietz05} where a change in $\chi$ in this
case was estimated by resumming the RPA series with the $g_4$ term.
However, in the presence of curvature, the RPA series does not
exhaust all possible processes, as the multi-tail fermionic loops
do not vanish, Eq. (\ref{triangle-full}). Hence the result of simple RPA
resummation is different from the answer by bosonization,
which implies the multi-tail fermionic processes, particularly, the
"drag diagrams", Fig.\ \ref{fig:drag1st}.

\subsection{Curvature in Luttinger liquid}
\label{rhorhoLutt}

Let us discuss now the modification of the theory in the presence of
the mixing interaction $\sim g_2 RL$  in (\ref{onewireHam}). In this
case right and left chiral densities mix, but the quadratic Hamiltonian
which is the
Luttinger model is still solvable. \cite{GoNeTs} The corresponding $u-v$
Bogoliubov transformation diagonalizes the Hamiltonian in terms of new
chiral densities, $\tilde R, \tilde L$, which obey the same commutation
relations (\ref{KacMoody}). Performing this transformation, one obtains

    \begin{eqnarray}
     { H}_0 &=& \pi v
     \left( \tilde R^2 + \tilde L^2\right),
         \label{onewireHam2} \\
     H_{cur} &=&
     \frac{\alpha_1}{3} (\tilde R^3 + \tilde L^3)
     + \frac{\alpha_2}{2}
     (\tilde R^2 \tilde L + \tilde R \tilde L^2) ,
     \label{curvatInt}
     \\
     \alpha_1 &=&
     \frac{\pi^2}{m}
     \frac{3+K^2}{2\sqrt{K}}
     , \quad
     \alpha_2 = \frac{\pi^2}{m} \frac{(K^2-1)}{\sqrt{K}},
    \end{eqnarray}
with  $v^2 = v_F(v_F + U/\pi)$, $K^2 = (1 + U/\pi v_{F})^{-1}$.
One can see from (\ref{curvatInt}) that apart from some rescaling of the
old mass term, $\tilde R^3 + \tilde L^3$, the effect of interaction
gives rise to a new type of the curvature vertex, which mixes new right
and left densities.

The density correlation function $\langle \rho \rho \rangle_{\omega q}$
is determined through the new variables $\tilde R, \tilde L$ as
\[ \langle \rho \rho \rangle_{\omega q}  =
 K\langle(\tilde R + \tilde L) (\tilde R + \tilde L)\rangle_{\omega q}.
\]
Note that $RL$ mixing, which was removed at the level of quadratic
action, reappears due to the curvature.
The mixed terms $\langle \tilde R
\tilde L\rangle$ are non zero in case of both  $U\neq 0$ and $m^{-1}
\neq0$, and it necessitates the consideration of the
(retarded) matrix Green function
 \[ \tilde G(\omega,q) =  \begin{pmatrix}
 \langle \tilde R \tilde R\rangle_{\omega q} ,&
 \langle \tilde R \tilde L\rangle_{\omega q} \\
 \langle \tilde L \tilde R\rangle_{\omega q}  ,&
  \langle \tilde L \tilde L\rangle_{\omega q}
 \end{pmatrix} \]
It is clear that the PT for bosons in the case $K\neq 1$ contains
processes absent in the free fermion case $K=1$.

Let us calculate the boson self-energy in the second order of PT.
Denoting three types of boson loops as $A_{RR}, A_{RL}, A_{LL}$ (Fig.),
one obtains
    \begin{eqnarray}
     \Sigma_{RR} &=&
     2\alpha_1^2 A_{RR} +\frac12 \alpha_2^2 A_{LL} + \alpha_2^2 A_{RL},
      \\
     \Sigma_{LL} &=&
     2\alpha_1^2 A_{LL} +\frac12\alpha_2^2 A_{RR} +\alpha_2^2 A_{RL} ,
     \\
     \Sigma_{RL} &=& \Sigma_{LR} =
     \alpha_1\alpha_2( A_{RR}+ A_{LL}) +\alpha_2^2 A_{RL}.
    \end{eqnarray}
Off the light cones, $\omega \neq \pm vq$, one has ${\rm Im}\; A_{RR} =
{\rm Im}\; A_{LL} = 0$ and all components of the imaginary part of the
self
energy are given by the quantity
        \begin{equation}
        \Sigma_{off}\equiv \alpha_2^2 A_{RL}.
        \end{equation}

We list here the form of the appearing (retarded) expressions:
    \begin{eqnarray}
    A_{RR} &=&\frac{1}{48\pi^3}
     \frac {q(q^2+ (2\pi T/v)^2)}{qv - \omega-i0},
     \label{ARR}
      \\
     A_{LL} &=&\frac{1}{48\pi^3}
      \frac {q(q^2+ (2\pi T/v)^2)}{qv + \omega+i0},
     \\
     {\rm Im}\; A_{RL} &=&
       \frac{\pi}{2} \frac{\omega^2-v^2q^2}{(4\pi v)^3}
       \left[ \coth\left(\frac{\omega+vq}{4T}\right)
       \right. \nonumber \\ && \left.
       +\coth\left(\frac{\omega-vq}{4T}\right)\right].
     \label{ARL}
    \end{eqnarray}
Note that the anomalous (appearing for $K\neq1$ only) contribution
$A_{RL}$ is ultra-violet divergent, which will be reflected in the
appearance of UV cutoff $\Lambda \sim E_F$ in real part of $A_{RL}$.
Seeking a function $A_{RL}$, analytical in the upper
semiplane of $\omega$, one can restore ${\rm Re}\; A_{RL}$ using the
rule
\[ i \coth \beta \omega \to - \frac{2}{\pi}\psi\left( - i\frac{\beta
\omega}{\pi}\right) + \frac1{i\beta\omega}.\]
Such a recipe of analytical continuation delivers ${\rm Re}\; A_{RL}$ up
to a
constant $\sim\ln(T/\Lambda)$ multiplied by $(\omega^2-v^2q^2)$, as will
be seen shortly.

In the limit of $T=0$, the expression for $A_{RL}$ is simplified
    \begin{eqnarray}
     {\rm Im}\; A_{RL} &=&
       \pi
       \frac{\omega^2-v^2q^2}{(4\pi v)^3}
      \Theta (\omega^2-v^2q^2)\, sign \,\omega,
      \\
      {\rm Re}\; A_{RL} &=&
        -\frac{1}{(4\pi v)^3}(\omega^2-v^2q^2)
       \ln\left|\frac{q^2v^2 - \omega^2}{\Lambda^2}\right|.
     \label{ARL0}
    \end{eqnarray}
Importantly, this part of $\Sigma$ vanishes at the light cone for $T=0$
and remains finite (and purely imaginary) for $T\neq0$, in contrast to
the singular contributions $A_{RR}, A_{LL}$. Notice the appearance of UV
cutoff $\Lambda$ in Eq.\ (\ref{ARL0}), which replaces temperature in
this limit.

Neglecting for a moment the singular (purely real) parts $A_{RR},
A_{LL}$, we write the Dyson equation in the form
    \begin{eqnarray}
     \tilde G^{-1} &=&
      \begin{pmatrix}
       \tilde g^{-1}_R -\Sigma_{off} ,&
       -\Sigma_{off} \\
       -\Sigma_{off} ,&
       \tilde g^{-1}_L -\Sigma_{off} \\
      \end{pmatrix},
    \end{eqnarray}
with $g^{-1}_{R,L}$  given by (\ref{bare-g}) with the change $v_F \to
v$. Some calculation shows then that the Green's function defined
for the initial fields $\phi, \theta$ is given by
    \begin{equation}
      \begin{pmatrix}
 \langle \partial \phi ,\partial \phi \rangle ,&
 \langle  \partial \phi,\partial \theta\rangle \\
 \langle\partial \theta ,\partial \phi \rangle , &
  \langle \partial \theta,\partial \theta\rangle
 \end{pmatrix}
      = \frac{\pi q^2}{Det}
      \begin{pmatrix}
       K v ,& \omega/q \\
        \omega/ q ,&
       K^{-1}(v -\Sigma_{off}/\pi) \\
      \end{pmatrix}  ,
      \label{Green1wire}
    \end{equation}
with
\[ Det = q^2 v^2 - \omega^2 - \frac{q^2v}{\pi} \Sigma_{off} ,\]
and $\partial$ standing for spatial derivative in (\ref{Green1wire}).

As a result, the density correlations are given by
    \begin{eqnarray}
      \langle\rho \rho\rangle_{\omega q} &=&
      \frac1{\pi^2}
     \langle \partial \phi \partial \phi \rangle_{\omega q}
      =  \frac{Kv q^2/\pi}{q^2v^2 - \omega^2 - q^2v
     \Sigma_{off}/\pi},
       \nonumber
     \\ &=&
     \frac{v_F}{\pi} \frac{q^2 Z}{q^2v^2 - \omega^2},
          \label{rhorho1}
     \\
    Z^{-1} &=& 1- \frac{q^2v\Sigma_{off}}{\pi(q^2v^2 - \omega^2)}.
    \end{eqnarray}
In the limit $T=0$ we have
    \begin{eqnarray}
     Z^{-1} &=&
     1 - \gamma (q/k_F)^2 \ln\frac{v^2q^2 -
      (\omega+i0)^2} {\Lambda^{2}},
     \\
     \gamma &=& \frac{K(K^2-1)^2}{ 64},
     \label{gamma}
    \end{eqnarray}
and $k_F = m v_F = Kmv$. The obtained logarithmic correction to the
residue $Z$ should be regarded as a first term in a series in $m^{-1}$.
This correction is increasingly important at the energies close to
light-cone condition, $\omega \to v|q|$, and corresponds to the
first term in the expansion of power-law singularity at $\omega \to
v|q|$, which is discussed in \cite{PKKG06}.

Away from the light cone, when the real part of logarithm in $Z$ can be
neglected, we obtain the contribution to the spectral weight at
$\omega>0$:
    \begin{eqnarray}
     {\rm Im}\; \langle\rho \rho\rangle_{\omega q} &\simeq&
     \frac{ \gamma}{k_F^2} \frac{v_F q^4 }{\omega^2-q^2v^2}
     \Theta (\omega^2-v^2q^2),
     \label{tailsT=0}
    \end{eqnarray}
which shows a tail at higher energies, $|\omega| >|qv|$.
This result was obtained earlier in \cite{PMGA,Teber06,G7}.

Finite temperatures, $T\agt |\omega|,|qv|$,
lead to a following modification :
    \begin{eqnarray}
     \frac{q^2v}{\pi}\Sigma_{off} &\simeq&
     i \gamma \frac{q}{k_F^2}4\pi\omega T,
    \end{eqnarray}
and from (\ref{rhorho1}) it appears that the bosonic excitations are
characterized by damping
    \begin{equation}
    \omega \simeq \pm vq + 2i\pi \gamma T (q/k_F)^2.
    \label{damping}
    \end{equation}
However, this ``damping'' should not be viewed as the characteristic
linewidth, because it is smaller than the previous estimate for the
width of the spectral width of density correlator in the noniteracting
case, $|\omega - vq| \sim T q/k_F$. This wider spectral width comes from
the omitted terms $A_{RR}, A_{LL}$ which intervene close to the light
cone condition $\omega =\pm vq$. The finite-$T$ ``damping'' due to
interaction manifests itself in the enhanced amplitude of the tails in
the density correlator, and instead of (\ref{tailsT=0}) we obtain
    \begin{eqnarray}
     {\rm Im}\; \langle\rho \rho\rangle_{\omega q} &\simeq&
     \frac{ \gamma}{2k_F^2}
     \frac{v_F q^4}{(\omega^2-q^2v^2)}
      \left[ \coth\left(\frac{\omega+vq}{4T}\right)
       \right. \nonumber \\ && \left.
       +\coth\left(\frac{\omega-vq}{4T}\right)\right],
     \\ &\simeq&
     \frac{ \gamma}{k_F^2} \frac{4v_F q^4 \omega T}
     {(\omega^2-q^2v^2)^2},
     \label{tailsTneq0}
    \end{eqnarray}
which is parametrically larger than (\ref{tailsT=0}) at
$|\omega -vq| \ll \omega \sim T $, and is
non-zero also at $|\omega| < |qv|$. It should be stressed again, that
Eqs.\ (\ref{tailsT=0}), (\ref{tailsTneq0}) are applicable only for
$|\omega - vq| \agt k_F^{-1} |q|\max(T, qv )$.

\section{a.c.\ conductivity of single wire}
\label{sec:conduc1}

Let us discuss the coupling to electromagnetic field in bosonization and
its modification due to curvature. We start from
general expression for the kinetic term in the Hamiltonian,
   \begin{equation}
   {\cal H}_{kin} = \frac 1{2m}\psi^\dagger (-i\nabla - eA)^2 \psi
   - \mu \psi^\dagger \psi,
   \label{kinetic0}
   \end{equation}
with $\hbar = c = 1$ and $\mu$ chemical potential. Using the above
representation $\psi(x) \sim e^{ik_F x +i\phi_R} +
e^{-ik_F x - i\phi_L}$, and denoting
\[  \Phi_R = k_Fx + \phi_R , \quad  \Phi_L = k_Fx + \phi_L \]
one can represent ${\cal H}_{kin}$ in the presence of $eA$ as
   \begin{eqnarray}
   {\cal H}_{kin} &=&
   \frac{1}{2\pi}
   \left[\frac{(\partial_x \Phi_R - eA)^3}{6m}
    - \mu \partial_x \Phi_R\right]
   \label{kinetic} \\ &&
    +    \frac{1}{2\pi}
   \left[\frac{(\partial_x \Phi_L + eA)^3}{6m}
    - \mu \partial_x \Phi_L\right].
    \nonumber
   \end{eqnarray}

Slightly digressing here, we note that in this notation $k_F
x$ is clearly a zero (classical) mode of the bosonic fields
$\Phi_{R,L}$ which is obtained from (\ref{kinetic}) by variation over
these fields. Note that the zero mode $k_Fx$ corresponds to a local
minimum of the action corresponding to (\ref{kinetic}), while generally
speaking the action is unstable with respect to choice $\partial_x
\Phi_{R,L} \to -\infty$. The latter unphysical choice of "classical"
vacuum can be ruled out by imposing the hard-core condition,  $
\partial_x \Phi_{R,L} > 0 $ for all $x$. It means that the
local fluctuations of the density should not lead to negative values of
total electronic density, $ \partial_x \phi_R > - k_F$ etc.

Note, that we can obtain (\ref{kinetic}) also
by including vector potential in the
$U(1)$ phase according to $\psi(x) \sim e^{i\Phi_R-ie\int dx\, A} +
e^{-i\Phi_L-ie\int dx\, A}$. The cubic terms in $eA$ generated by two
terms in (\ref{kinetic}) cancel each other.
In more general case of non-parabolic band dispersion,
$\varepsilon(k)$, the Peierls substitution, $k \to k - eA$,
leads to higher powers of $eA$ in the gradient expansion of
$\varepsilon(k-eA)$. The recipe for
tackling this case was outlined long ago by Luttinger and Kohn
\cite{LuttKohn55} and amounts to retaining only $eA$ and $(eA)^2$
terms in such expansion, while discarding higher terms as unphysical.

Upon thermodynamic averaging of ${\cal H}_{kin}$, the
diamagnetic term, $\sim (eA)^2$, gives the usual value, $k_F/(\pi m) =
Kv/\pi$. This is the zero-mode contribution, whereas terms linear in
$\partial_x \phi$ yield zero.
The term linear in vector potential has a structure
      \begin{eqnarray*}
    && \frac{eA}{2m}[ ( k_F + \partial_x \phi_L )^2 -
   ( k_F + \partial_x \phi_R )^2 ]
    \\ &=&
   \pi^{-1} eA( v_F \partial_x \theta + m^{-1}
   \partial_x\theta \partial_x\phi)
   \end{eqnarray*}
The last expression means that the electric current is given by the
expression
   \begin{equation}
    j =  - e\pi^{-1}(v_F \partial_x \theta +   m^{-1}
   \partial_x\theta \partial_x\phi).
   \label{current}
   \end{equation}
This expression is confirmed by the immediate application of the
continuity equation, $\partial_x j = - \partial_t \rho = ie
\pi^{-1}[\partial_x\phi, {\cal H}]$, and using the bosonized Hamiltonian
(\ref{Ham1wire}). The first term in (\ref{current}) is usual for LL, the
second term appears due to fluctuation of the Fermi velocity when
changing the density.

The Kubo formula expresses the conductivity via the retarded Green's
function of the uniform current,
   \begin{equation}
   \sigma(\omega) =
   \frac{\langle j j \rangle_{q=0,\omega} -
   \langle j j \rangle_{q=0,\omega=0}
   }{i\omega}.
   \label{Kubo}
   \end{equation}
The subtraction of the diamagnetic term $\langle jj
\rangle_{q=0,\omega=0}$ should be performed when using
direct representation (\ref{current}). If one calculates first the
density correlator and uses the continuity equation afterwards,
\[
\langle jj \rangle_{q=0,\omega} = \lim_{q\to0}
\omega^2q^{-2}
\langle \rho \rho \rangle_{q,\omega},
\]
then the diamagnetic contribution is subtracted automatically.

Using the above result, Eq.\ (\ref{rhorho1}), we have
   \begin{equation}
   \sigma(\omega) =
   \frac{v_F}{-i\pi \omega}, \quad
   {\rm Re}\; \sigma(\omega) = v_F \delta(\omega),
   \label{conduc1wire}
   \end{equation}
i.e.\ the conductivity of {\em one wire} is not affected by the
temperature, interaction and curvature, in accordance with Giamarchi and
Millis. \cite{GiaMil92} It should be stressed that this conclusion
assumes our above choice $g_2 = g_4 =  U$, i.e.\ physical charge-charge
interaction ; we do not consider models with $g_2 \neq
g_4$ which would correspond to current-current interaction and
to modifications of (\ref{conduc1wire}).

We also comment here on the equivalence of two ways of calculation of
$\sigma(\omega)$, via Eq.\ (\ref{current}) and via continuity
equation. Using (\ref{current}) for $\langle jj\rangle$ correlation
function, one obtains a bosonic propagator $\langle
\partial_x \theta, \partial_x \theta \rangle$, and bosonic loops of
the form $m^{-2}\langle
\partial_x \theta \partial_x \phi ,\partial_x \theta \partial_x
\phi\rangle$. From Eq.\ (\ref{Green1wire}) the average
$\langle \partial_x \theta, \partial_x \theta \rangle$ is vanishing in
the limit $q \to 0$, and subtracting the diamagnetic contribution
$v_F/\pi$, one arrives at (\ref{conduc1wire}). The bosonic loop terms
give zero, as will be seen shortly.

\begin{figure}
\includegraphics[width=4cm]{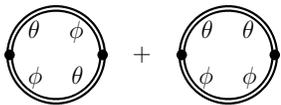} %
\caption{Two boson loop diagrams formally contributing to a.c.\
conductivity. The labels $\theta \phi$, etc., at boson lines refer to
the boson fields as disussed in text.
\label{fig:dragloop}}
\end{figure}

\begin{figure}
\includegraphics[width=8cm]{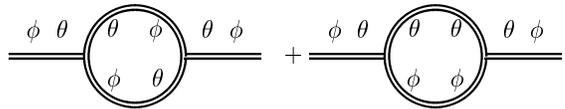} %
\caption{Two boson diagrams for  a.c.\ conductivity,
obtained with the use of continuity relation and equivalent to Fig.\
\ref{fig:dragloop}.
\label{fig:123}}
\end{figure}

The mixed terms in $\langle jj\rangle$ correlator,
$m^{-1}\langle
\partial_x \theta, \partial_x \theta \partial_x \phi \rangle$ are
zero in all orders of perturbation theory. It is trivial to see in
lowest order of $m^{-1}$, because the mixed average contains three
bosons. In the next order, one has a diagram with a bosonic loop and the
Greens's function attached to it. This Green's function bears the zero
external momentum and finite $\omega$ and hence vanishes.
The same argument about zero contribution to conductivity can be applied
to any diagram containing external lines carrying zero momentum,
particularly in the two-wire situation. Hence, apart from the
$\delta-$function contribution produced by conventional bosonization,
we have to discuss only loop diagrams in Fig.\ \ref{fig:dragloop} and
their dressing by the interaction, $m^{-1}$.

\section{a.c.\ conductivity, two wires}
\label{sec:conduc2}

For the system described by the Hamiltonian (\ref{ham}), we wish to
compute the transresistivity, i.e. the voltage drop in one wire, caused
by the current in another wire. Technically, we use Kubo
formula and calculate the conductivity, not resistivity. We consider
the drag effect which is caused purely by electronic interaction, and
in the absence of disorder in wires. In this situation, one expects a
$\delta-$ function contribution to the conductivity, which should be
separated from the proper drag contribution, seen in the limit
$\omega\to0$. Let us first discuss this small-$\omega$ limit.

\subsection{d.c. limit}

A general form of transconductivity in the limit $\omega \to 0$ can be
found, when starting with the transresistivity matrix $\hat R
(\omega) =  [\hat \sigma(\omega) ]^{-1} $. Exactly at $\omega = 0$  the
voltage drop in both wires should be absent when the drift charge
velocities, $\pi j_i/k_{Fi}$, in both wires coincide. In order to see it
suffices to make a Galilean  transformation to the coordinate frame
moving with the drift velocity. \cite{PMGA} We combine this
observation with the third Newton law, equalling the forces acting on
both wires, i.e. $E_1 k_{F1}=-E_2 k_{F2}$, and represent the
complex-valued impedance as a series in $\omega$ :
        \begin{eqnarray}
       \hat R(\omega) &=&
        \pi r_0
       \begin{pmatrix} k_{F1}^{-2}, & -k_{F1}^{-1}k_{F2}^{-1} \\
       -k_{F1}^{-1}k_{F2}^{-1}, & k_{F2}^{-2}    \end{pmatrix}
         \label{resist1}
          \\ &&
        -i\omega \hat A + \omega^2 \hat A_2 + \ldots,\nonumber
       \end{eqnarray}
with the transresistivity given by $R_{12}  = \pi r_0 /(k_{F1} k_{F2})
$.

In the absence of interwire interaction $U_{12}$ in
(\ref{ham}), Eq.\ (\ref{conduc1wire}) shows that $r_0=0$ , $\hat A_2 =
0$ and $\hat A = \pi\, diag [v_{F1}^{-1}, v_{F2}^{-1}]$. We make a
natural
conjecture, that $r_0, \hat A, \hat A_2$ allow Taylor expansions
in $u$. It follows then, that for $\omega \alt r_0 v_J k_{F}^{-2}$ the
first two terms in (\ref{resist1}) are of the order of $r_0 k_{F}^{-2}$,
and the third term is $\sim (r_0 k_{F})^{-2} O(u)$, i.e.\ contains an
additional small factor $O(u r_0)$. Therefore one can invert $\hat
R(\omega)$ at small $\omega$, retaining only first two terms in
(\ref{resist1}) and taking the above non-perturbed value of  $\hat A$.

The resulting complex-valued conductivity is obtained as
      \begin{eqnarray}
      \hat \sigma(\omega) &=&
      \hat \sigma_1(\omega) + \hat \sigma_2(\omega),
      \label{conduc1} \\
       \hat \sigma_1(\omega) &=&
       \frac{ \pi^{-1}d^{-1} }{ +0 -i\omega}
         \begin{pmatrix}
       v_{F1} v_{F2}^{-1}m_2^{-2} ,& m_1^{-1}m_2^{-1} \\
      m_1^{-1}m_2^{-1} ,& v_{F1}^{-1} v_{F2}m_1^{-2}
        \end{pmatrix},
      \nonumber \\
       \hat \sigma_2(\omega) &=&
        \frac{\pi^{-1} d^{-1}}{r_0 d -  i\omega}
        \begin{pmatrix} m_1^{-2} ,& -m_1^{-1}m_2^{-1} \\
        - m_1^{-1}m_2^{-1} ,& m_2^{-2}    \end{pmatrix},
        \label{sigma2}
        \\
        d &=& {m_1^{-2} v_{F1}^{-1}} + {m_2^{-2} v_{F2}^{-1}}.
        \nonumber
       \end{eqnarray}
Interestingly, (\ref{conduc1}) provides the universal, i.e., independent
of details of fermionic interaction, relation between the drag
resistivity and the drag conductivity in small-$\omega$ limit,
        \begin{equation}
        R_{12} \, {\rm Re}\;[\sigma_{12}] \,\rvert_{\omega\to 0 } =
      (c+c^{-1})^{-2},
        \label{relationRsigma}
        \end{equation}
with $c^2 = m_1 k_{F1}/(m_2 k_{F2})$. The r.h.s. of
(\ref{relationRsigma}) is always less than $1/4$, which value is
achieved for equal wires.

In the continuum media, the integral of $\sigma(\omega)$ over all
frequencies should not depend on the interactions in the system, which
statement is known as the optical sum rule (see, e.g., \cite{AriZey05}
and references therein). From (\ref{conduc1}) we restore this statement
~:
        \begin{equation}
     {\rm Re}\; \int_{-O(r_0d)}^{O(r_0d)} d\omega \hat \sigma(\omega)
\simeq
    \pi \begin{pmatrix} v_{F1} ,& 0 \\
    0 ,& v_{F2} \end{pmatrix} ,
    \label{sumrule}
    \end{equation}
which means that the spectral weight of the part $\hat
\sigma_2(\omega)$ is large $\sim v_{F}$ and is concentrated at the
lowest energies. Regrettably, this is the worst accessible region for
bosonization analysis, as will be seen shortly.

We can cast (\ref{conduc1}) in the form
      \begin{eqnarray}
      \hat \sigma(\omega) &=&
         \frac{\pi^{-1}}{-i \omega}
        \begin{pmatrix}v_{F1} ,& 0 \\
        0 ,&  v_{F2}  \end{pmatrix}
        \label{sigmaExpan} \\ &&
        +
        \frac{\pi^{-1} r_0}{\omega(\omega + i r_0 d )}
        \begin{pmatrix} m_1^{-2} ,& -m_1^{-1}m_2^{-1} \\
        - m_1^{-1}m_2^{-1} ,& m_2^{-2}    \end{pmatrix},
        \nonumber
       \end{eqnarray}
and identify the first term here with the bare bosons contribution
(\ref{conduc1wire}), then the second term is entirely the effect of
the curvature. At larger $\omega$, we have the expansion
\[ {\rm Re}\; \sigma \sim \frac{r_0  m^{-2}}{\omega^2 + r_0^2 d^2}
\propto
 \frac{r_0 d}{\omega^2} - \frac{r_0^3  d^3 }{\omega^4} + \ldots
\]
The drag effect, $r_0 d \neq 0$, is absent without curvature,
and bosonization treats the curvature as perturbation.
It means, particularly, that calculating the d.c.\ drag
resistivity starting from Kubo formula, (\ref{Kubo}), one
should seek a lowest order contribution to ${\rm Re}\;  \sigma(\omega)$
behaving as $\omega^{-2}$. Such contribution should arise in
the higher-than-second order of $m^{-1}$, and the next important term,
$\omega^{-4}$, is found in even higher order of PT.
Our approach in determination of $r_0$ as the coefficient before the
$\omega^{-2}$ term in conductivity is thus similar to
memory function formalism. \cite{GoeWoe72}

Below we prove that apart from the singular $\omega = 0$ contribution,
the matrix $\hat \sigma(\omega)$ is given by
      \begin{eqnarray}
      \hat \sigma(\omega) &=&
      \sigma_{d}(\omega)
      \begin{pmatrix} m_1^{-2} & -m_1^{-1}m_2^{-1} \\
      - m_1^{-1}m_2^{-1} & m_2^{-2}    \end{pmatrix}.
      \label{sigmaGen}
      \end{eqnarray}
This degenerate matrix cannot be inverted and one needs the complete
representation (\ref{conduc1}) in order to obtain $R_{12}$. From now on
we discuss the real part of the drag conductivity $\sigma_{d}(\omega)$.

\subsection{Optical Transconductivity}

For the Hamiltonian (\ref{ham}), the bare Green's function defined for
the vector $\Phi = ( \partial_x\theta_1, \partial_x \phi_1,\partial_x
\theta_2,  \partial_x \phi_2)$ is found as
    \begin{eqnarray}
     G  &=& \pi
     \begin{pmatrix}
     v_{1J} ,& \omega/ q ,& 0,& 0 \\
     \omega /q ,& v_{1N} ,& 0,& U_{12}/\pi \\
     0,& 0,& v_{2J} ,& \omega /q \\
     0,& U_{12} /\pi ,& \omega /q ,& v_{2N}
     \end{pmatrix}^{-1}.
     \label{Green}
    \end{eqnarray}
It shows that the excitation spectrum has two branches
$\varepsilon_{+q} = v_+ |q| $ and $\varepsilon_{-q}= v_- |q| $
with the sound (plasmon) velocities
    \begin{eqnarray}
     v_\pm^2 &=&
     \frac 12 (v_1^2 + v_2^2) \pm
     \sqrt{ \frac 14 (v_1^2 - v_2^2)^2 + u^2 v_1^2  v_2^2},
     \label{velocities}
    \end{eqnarray}
we assume $v_+ > v_- >0$ below.
We also focus on the case of almost identical wires, $v_1 \simeq
v_2 = v$ and small $u$, when $v_+ \simeq v_- \simeq v$ and
\begin{equation}
 \Delta \equiv (v_+-v_-)/v_- \simeq \sqrt{(v_1-v_2)^2/v^2 + u^2} \ll 1 .
 \label{defDelta}
\end{equation}

Let us discuss the first diagrams appearing in bosonization analysis of
transconductivities. As shown above, the linear-in-bosons components of
the currents (\ref{current}), $\partial\theta_i$, produce the
$\delta-$function contribution to the conductivity. This is the first
term in Eq.\ (\ref{sigmaExpan}).

The proper transconductivity part, $\sigma_d(\omega)$ in
(\ref{sigmaGen}) is generated by the second term of the current
operator in (\ref{current}). Schematically, it is given by two
diagrams, depicted in Fig.\ \ref{fig:dragloop}.
These bosonization diagrams correspond to the usual
fermionic ones,  Fig.\ \ref{fig:drag1st}, after we identify
Fig.\ \ref{fig:drag1st} with Fig.\ \ref{fig:123} in the approach
using the continuity relation, and notice that
the external tail Green's function become constant in the limit $q\to
0$, $\omega\neq0$.

After a lengthy calculation, sketched in Appendix, we obtain
    \begin{eqnarray}
     \sigma_d(\omega) &= & u^2
    \frac{\omega \sinh(\omega/2T)
    \; v_1^2 v_2^2  v_+^{-1} v_-^{-1}
     (v_+  +v_-)^{-5} }{8
    \sinh(\frac{v_+\omega}{2T (v_++v_-)})
    \sinh(\frac{v_-\omega}{2T (v_++v_-)})    }
             \nonumber \\ &&
     + (v_- \to -v_-),
     \label{opt.condu.finiteT}
     \\
     \sigma_d(0) &= &
     T  \left(\frac {u v_1 v_2 }{ v_+  v_-}\right)^2
      \frac{v_+^3  + 3v_+ v_-^2 }{2(v_+^2 -v_-^2)^3}.
     \label{condu0}
    \end{eqnarray}
First observation about the obtained optical conductivity is that
$\sigma_d(\omega)$
non-zero solely due to the interwire interaction $u$.
Secondly, at $\omega \agt T$ the second term in
(\ref{opt.condu.finiteT}) is  exponentially
small and the first one produces linear-in-$\omega$ optical
conductivity,
    \begin{equation}
    \sigma_d (\omega) \sim u^2 v_F \omega / E_F^2,
    \quad \omega \agt T      .
    \end{equation}
Unusually, there is no upper boundary for this regime in the
theory, and $Re \sigma_d (\omega) \propto |\omega|$  until
$|\omega| \alt E_F$. At the same time,
the integral contribution of higher frequencies into the optical
sum is small, $\int d\omega\,  \sigma(\omega) \sim u^2 v_F$, i.e.
only a fraction of the above value (\ref{sumrule}).

In the case of identical wires, $v_1 = v_2 \equiv v$, coupled by a weak
interaction, one has $u\ll 1$ and the new velocities  $v_{\pm} = v
\sqrt{1\pm u}$. The optical conductivity takes then the form
     \begin{equation}
     \sigma_d(\omega) \simeq
     \frac{u^2 \omega \coth(\omega/4T)   }{128 v^3} +
     \frac{\omega \sinh(\omega/2T) }{8 u^3 v^3 \sinh^2(\omega/2uT)}.
     \label{cond.ident.wires}
     \end{equation}
One sees the appearance of a smaller energy scale $uT \ll T$ in this
case and parametric enhancement $\sigma_d(0) \sim T/u$, we return to
this point below.

Overall, the function (\ref{opt.condu.finiteT}) is a smooth function of
$\omega$,  not exhibiting the internal crossover at $\omega \sim r_0d
\sim  m^{-4}$ as suggested by (\ref{sigmaExpan}). The region of
applicability of the expression (\ref{opt.condu.finiteT}) can be
estimated by using the "broadened" propagators (\ref{exacpropT})
instead of $\delta-$function-like (\ref{bare-g}), during calculation of
$\sigma_d(\omega)$. This estimate shows that the finite bosons'
linewidth intervenes the formula (\ref{opt.condu.finiteT}) at $\omega
\alt \omega_0 = T^2/mv^2$, in accordance with \cite{Samo98}. Using this
estimate as a crossover energy from (\ref{opt.condu.finiteT}) to the
Lorentzian form (\ref{sigmaExpan}), we can write
$\sigma_d(\omega_0) \simeq r_0/\omega_0^2$ and thus find $r_0$ up to a
numerical coefficient. Remarkably, this simple method gives a correct
value of $r_0$ at lowest $T$, but it wrongly estimates a crossover to
high-temperature regime, discussed below. We see that the estimate of
$r_0$ requires going beyond the lowest order of curvature and we suggest
a way to calculate it in the next subsection.

\subsection{Alternative representation of $\sigma_d(\omega)$}

In this subsection we derive another representation for
$\sigma_d(\omega)$, which explicitly recovers the matrix structure
(\ref{sigmaGen}) and shows the overall prefactor $u^2$ in
$\sigma_d(\omega)$. For infinitesimal $u$ this representation yields
the well-known formula for the drag coefficient.

We rewrite the Kubo formula as
   \begin{equation}
   \sigma(\omega) =
   \frac{\langle \partial_t j  \partial_t j \rangle_{q=0,\omega}
    }{i\omega^3}.
   \label{Kubo2}
   \end{equation}
and use the equation of motion  $\partial_t j  = i [{\cal H},j]$ and
Eq.\ (\ref{current}). Some lengthy, but straightforward calculation
verifies that the commutation with the terms $H_1,
H_2, H_{cur}$ produces full derivatives with respect to $x$, which
vanish for the uniform currents. The only surviving terms come from
$H_{12}$ in (\ref{ham}) and are
   \begin{eqnarray}
    \partial_t j_1(x) &=&U_{12} m_1^{-1} \rho_1 \partial_x \rho_2,
    \\
    \partial_t j_2(x) &=& U_{12} m_2^{-1} \rho_2 \partial_x \rho_1 \sim
    -U_{12}m_2^{-1}\rho_1 \partial_x \rho_2     ,
   \end{eqnarray}
modulo full derivatives. It is immediately seen that the structure
(\ref{sigmaGen}) is
reproduced, and the scalar drag conductivity is given by
   \begin{eqnarray}
    \sigma_d(\omega) &=&
    \frac{U_{12}^2 }{i\omega^3}
    \langle
    \rho_1 \partial_x \rho_2, \rho_1 \partial_x \rho_2
    \rangle_{q=0,\omega}.
    \label{Kubo3}
   \end{eqnarray}
Comparing  to (\ref{sigmaExpan}), we obtain at $\omega >  r_0 d$
   \begin{eqnarray}
    r_0 &=&
    \frac{U_{12}^2 }{\omega} {\rm Im}\;     \langle
    \rho_1 \partial_x \rho_2, \rho_1 \partial_x \rho_2
    \rangle_{q=0,\omega}.
    \label{r0-gen}
   \end{eqnarray}

For the {\em infinitesimal} $U_{12}$ the cross-terms,
$\sim \langle \rho_1 ,\partial_x \rho_2  \rangle$, in
the average (\ref{Kubo3}) can be neglected, and a comparison with Eq.\
(\ref{sigmaExpan}) at $\omega \to 0$ gives
   \begin{eqnarray}
    r_0 &=&
    \pi U_{12}^2 \int \frac{q^2dq\,d\omega}{(2\pi)^2}
    \frac{ {\rm Im}\;\chi_1(q,\omega){\rm Im}\;\chi_2(q,\omega)}
    {2T\sinh^2(\omega/2T)},
    \label{dragcoef}
   \end{eqnarray}
which corresponds to previous findings. \cite{PMGA,drag90s} It is also
worth noting that our Eq.\ (\ref{r0-gen}) for transresistivity exactly
corresponds to Eq.\ (18) in \cite{AmiVig00}. We should stress, however,
that the {\em exact formula} for the case of clean Luttinger liquids is
given by Eq.\ (\ref{Kubo3}), and (\ref{r0-gen}) should be regarded as
the high-energy asymptote of the expression (\ref{sigmaExpan}).

\subsection{Non-equal wires}

We consider now the case of two wires, which are coupled by
infinitesimal $U_{12}$ and are characterized by non-equal
plasmon velocites. In this case we can use Eq.\
(\ref{dragcoef}) and the results of the Section \ref{rhorhoLutt}.
In the absence of curvature and intrawire interaction, the formula
(\ref{rhorho1}) with $Z=1$ gives $r_0 = 0$ in (\ref{dragcoef}). The
first non-vanishing contribution is given by (\ref{tailsTneq0}),
and we obtain
        \begin{eqnarray}
    r_0 &=&
    \frac{ U_{12}^2 v_{F1} v_{F2} }{16 \pi T (v_1^2 - v_2^2)}
    \int \frac{q^5dq}{\sinh \frac{(v_1-v_2) q}{4T}
    \sinh \frac{(v_1+v_2)q}{4T} }
    \nonumber \\  &\times&
    \left[
    \frac{\gamma_2}{v_1 k_{F2}^2 \sinh(v_1 q/2T)}
    + (1\leftrightarrow 2)
    \right].
    \label{dragcoef1a}
   \end{eqnarray}
In the important case of almost identical wires, when $\Delta =
(v_1-v_2)/v \ll 1$, we can omit the wire index in the prefactors,
expand  $ \sinh ({(v_1-v_2) q}/{4T}) \simeq \Delta v q/4T$
and get
    \begin{eqnarray}
    r_0      &\simeq&
     \frac{8\pi^3 U_{12}^2}{m^2  v^4 \Delta^2}
    \frac{\gamma T^5}{15 v^5},
    \label{dragcoef1b}
   \end{eqnarray}
with $\gamma$ given by (\ref{gamma}) and we used
$\int_{-\infty}^\infty dx\, x^4/\sinh^2 x = \pi^4/15$.
Comparing (\ref{dragcoef1b}) to the result by Pustilnik {\em et al.},
\cite{PMGA}
we see that we obtained the same $T^5$ dependence but parametrically
larger prefactor, $(v_1-v_2)^{-2}$, rather than $(v_1-v_2)^{-1}$.

Note that one would reproduce the prefactor obtained in \cite{PMGA},
if one neglected the $T-$dependence of density propagator and
substituted in (\ref{dragcoef}) the expression (\ref{tailsT=0}) for
${\rm Im}\;\chi_1(q,\omega)$ calculated at $T=0$, instead of
(\ref{tailsTneq0}).
We see that finite $T$ parametrically enhances
the ``tail'' of density correlator, and it has consequences for the
drag effect. The importance of calculation of intermediate polarization
operators ${\rm Im}\;\chi_i(q,\omega)$ in (\ref{dragcoef}) at finite
$T$ was also emphasized in \cite{GoMivOp04}.

Checking the applicability of (\ref{tailsTneq0}), one can conclude that
the derivation of (\ref{dragcoef1b}) assumed a condition $v_1-v_2 \agt
T/k_F$. In other words, the expression (\ref{dragcoef1b}) should be
regarded as \textit{low-temperature} regime of drag resistivity, taking
place at
 \[ T \alt T_1 \equiv k_F(v_1-v_2) = E_F \Delta .\]
Above this temperature scale, the "core" spectral
weights of $\chi''_1(q\omega)$ and $\chi''_2(q\omega)$ overlap
and one can estimate the
drag coefficient, by taking the expression for the "core" of the
density propagator for free fermions, Eq.\ (\ref{exacpropT}), to obtain
  \begin{equation}
   r_0  \sim \frac{U_{12}^2 m}{v^3} T^2
   F\left(\frac{T_1}{2T} \right).
   \label{simpleT2}
  \end{equation}
with
    \begin{eqnarray}
   F(z) &=& \int_0^\infty
   \frac{2x\,dx (x\coth x - z\coth z)}{\sinh(x+z)\sinh(x-z)} ,
   \nonumber \\
   &=& 1 , \quad z \to 0 , \nonumber \\
   &\simeq& 4z^3 e^{-2z}/3,\quad z \gg 1 .\nonumber
  \end{eqnarray}
The estimate (\ref{simpleT2}) shows an activational behavior of the drag
for free fermions, similarly to Ref.\ [\onlinecite{PMGA}].

The obtained estimates for the drag effect at lower and higher
temperatures,  however, raise a few questions.
First, the Eqs.\ (\ref{simpleT2}) and
(\ref{dragcoef1b}) apparently describe different contributions to the
drag,  they do not match at the suggested
crossover temperature  $T \sim T_1$, and Eq.\ (\ref{dragcoef1b}) is
explicitly non-zero only due to
intrawire interactions and this feature is absent in (\ref{simpleT2}).

Second, in case of initially equal wires,
the plasmon velocities are split due to interwire
interaction, Eq.(\ref{defDelta}). It means that $U_{12}$ cannot be
regarded as infinitesimal in (\ref{Kubo3}) and hence
Eq.\ (\ref{dragcoef1a}) is, generally speaking, invalid. In this case a
naive substitution $\Delta \sim U_{12}/ v$ from (\ref{defDelta}) results
in $r_0$, Eq.\ (\ref{dragcoef1b}) independent of interwire coupling.

To clarify these issues, we separately consider the case of nearly
equal wires below.

\section{drag for nearly equal wires in bosonization}
\label{sec:nearlyequal}

\subsection{Identical wires, Hamiltonian}

When the wires are identical, it is convenient to introduce the
symmetrized combinations
\begin{equation}
\sqrt{2} \phi_{\pm} = \phi_1 \pm \phi_2, \quad
\sqrt{2} \theta_{\pm} = \theta_1 \pm \theta_2.
\end{equation}
A technical convenience of $\sqrt{2}$ prefactor stems from the
same commutation relations for the symmetrized operators, however the
zero modes are different, $\partial_x \phi_{+}^{(0)} = \sqrt{2} k_F$,
$\partial_x \phi_{-}^{(0)}  =0$. We will call $\phi_{+}$ and $\phi_{-}$
pseudocharge and pseudospin fields, by obvious analogy with
bosonization for one wire in the spinful case. \cite{GoNeTs}

In terms of densities and momentum
operators,
    \begin{equation}
    \rho_\pm = \pi^{-1}\partial_x \phi_{\pm} , \quad
    \Pi_\pm = \pi^{-1}\partial_x \theta_\pm    ,
    \end{equation}
the  Hamiltonian
(\ref{ham}) is rewritten as follows
    \begin{eqnarray}
    {\cal H} &=& \int dx\,( H_+ + H_-  +
     H_{cur} ) ,
     \label{hamsymm} \\
     H_{\pm} &=&\frac{\pi}{2}(  v_{J} \Pi_\pm^2 + v_{N} (1\pm u)
    \rho_\pm^2) , \\
     H_{cur} &=&
     \pi^2\frac {
      \rho_+ \left(\rho_+^2 + 3 \Pi_+^2
     + 3 \Pi_-^2 + 3\rho_-^2 \right ) +
     6\Pi_+ \rho_- \Pi_- }
      {6m\sqrt{2}}  .
     \nonumber
    \end{eqnarray}
It is seen that the change in the zero mode of $\phi_{+}$ is
compensated by the enhancement of its ``mass''. The plasmon velocities
and Luttinger parameters are given by
   \begin{eqnarray}
   v_\pm^2 &=&   (1\pm u )v_{N} v_{J}  , \quad
   K_\pm^{-2} = (1\pm u )v_{N} / v_{J}.
   \end{eqnarray}
According to our definitions
   \[ K_\pm^{-2} = 1 + (\pi v_F)^{-1} (U_1 \pm U_{12}),   \]
so that $K_- = 1$ for equal intrawire and interwire interactions.

We diagonalize the Hamiltonian (\ref{hamsymm}) by transformation
   \begin{eqnarray}
    \rho_\pm &=& K_\pm^{1/2}\tilde \rho_\pm ,
    \quad
    \Pi_\pm = K_\pm^{-1/2} \tilde \Pi_\pm .
   \end{eqnarray}
In terms of new left and right movers,
$2\tilde R_\pm =   \rho_\pm + \Pi_\pm $,
$2\tilde L_\pm =   \rho_\pm - \Pi_\pm $,
there are no mixed terms $\tilde R_\pm \tilde L_\pm$ in $\cal H$, and
propagators $\langle \tilde R_\pm \tilde L_\pm \rangle$ vanish.

The curvature term becomes
    \begin{eqnarray}
     H_{cur} &=&
     \frac{\alpha_1}{3} \left(
     \tilde R_+^3 + \tilde L_+^3\right) +
     \frac{\alpha_2}{2} \left(\tilde R_+^2 \tilde L_+
     + \tilde R_+ \tilde L_+^2\right)
     \nonumber \\  && +
     \frac{\alpha_3}2
     (\tilde R_+\tilde R_-^2 + \tilde L_+\tilde L_-^2)
     + \frac{\alpha_4}2
     (\tilde R_+\tilde L_-^2 + \tilde L_+\tilde R_-^2)
     \nonumber \\  &&
     + \alpha_5 ( \tilde R_+ + \tilde L_+ ) \tilde R_- \tilde L_-
     , \label{curvaEqual}
    \end{eqnarray}
with
    \begin{eqnarray}
    \alpha_{1} &=&
     \frac{\pi^2}{m}
     \frac{3+K_+^2}{2\sqrt{2K_+}}
     , \quad
     \alpha_{2} =
     \frac{\pi^2}{m\sqrt{2}} \frac{(K_+^2-1)}{\sqrt{K_+}},
     \nonumber \\
     \alpha_{3(4)} &=&
     \frac{\pi^2\sqrt{K_+}}{m\sqrt{2}}\left(
     \frac1{K_-} + K_-  \pm \frac2{K_+}    \right)
      ,\nonumber  \\
     \alpha_{5}  &=&
     \frac{\pi^2\sqrt{K_+}}{m\sqrt{2}}
     \left( K_- - \frac1{K_-}\right),
     \label{alphas}
    \end{eqnarray}
in the limit of vanishing interwire interaction we have $K_+ = K_-$
and $2\alpha_1 = \alpha_3$ and $\alpha_2 = \alpha_4 =\alpha_5$; this
case returns us back to Eq.\ (\ref{curvatInt}).
In the absence of interaction, $K_\pm =1 $ and $\alpha_{2} = \alpha_4
=\alpha_5 = 0$. Note that the vertex $\alpha_{1}$, describing the decay
within the same mode, is always non-zero.

\begin{figure}
\includegraphics[width=1in]{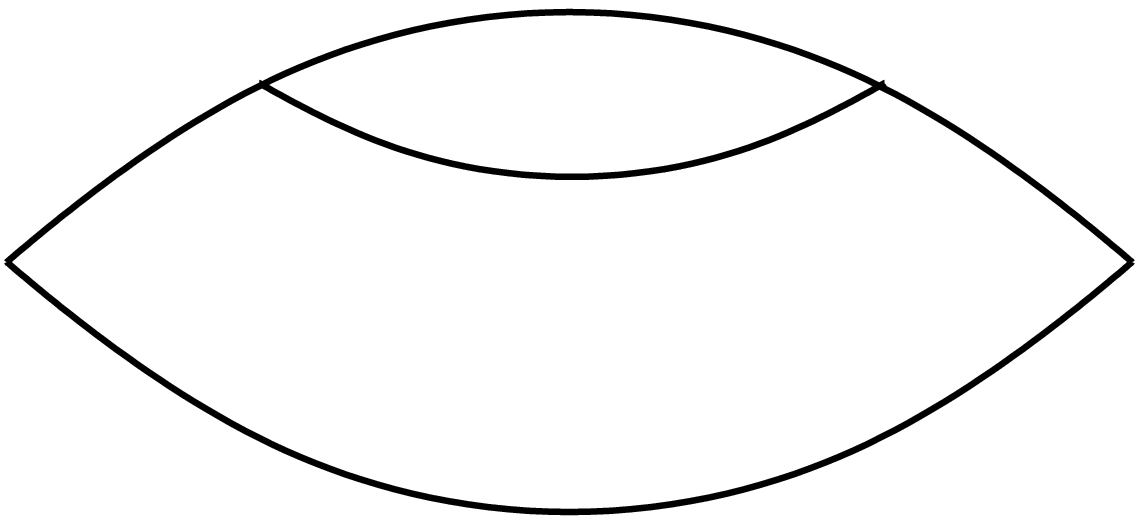}%
\hspace{.7cm}\includegraphics[width=1in]{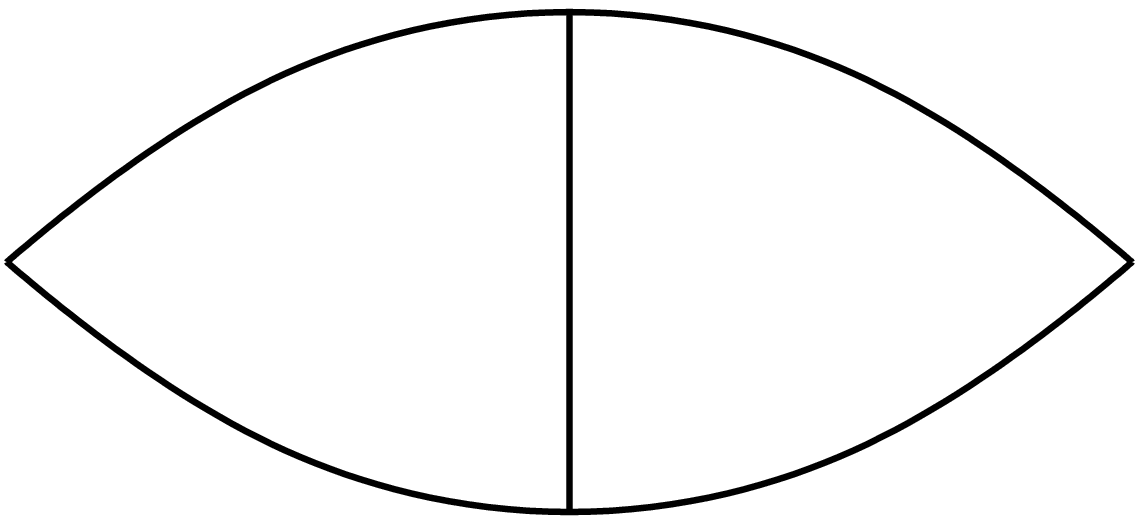}%
\caption{ Skeleton diagrams depicting the bosonic processes
contributing to drag effect at lowest temperatures.
\label{fig:fish1}}
\end{figure}

\begin{figure}
\includegraphics[width=0.4\columnwidth]{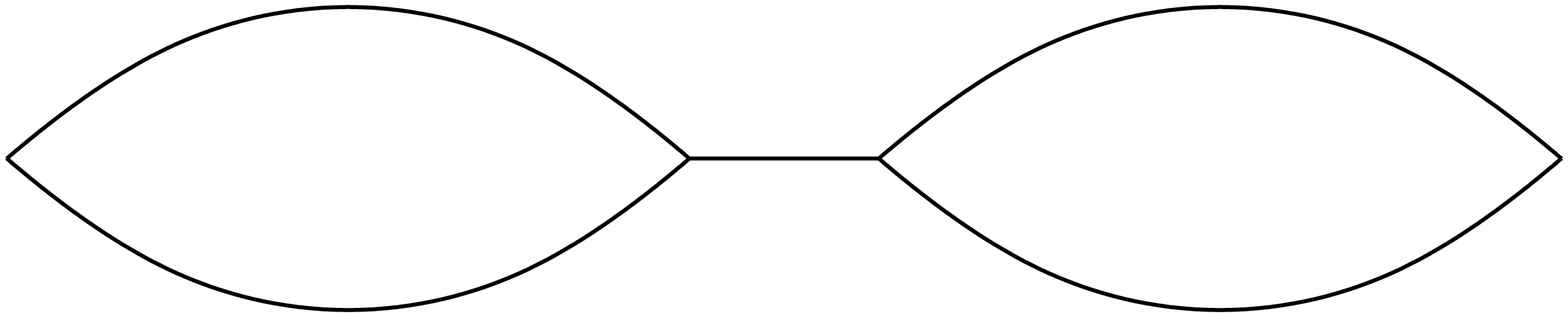}%
\caption{ RPA processes, which do not contribute to drag
conductivity. \label{fig:twofish}}
\end{figure}

The drag current operator in (\ref{Kubo3}) is
\begin{equation}
\rho_1 \partial_x \rho_2 = \rho_- \partial_x \rho_+
 = (K_+ K_-)^{1/2} \tilde \rho_- \partial_x \tilde \rho_+,
 \label{ext-vertex}
\end{equation}
modulo full derivatives. We seek the contribution to $r_0$ in the next
order in $m^{-1}$ which can be obtained by considering two skeleton
graphs, depicted in Fig.\ \ref{fig:fish1}.
The diagrams of the RPA type, Fig.\  \ref{fig:twofish}, vanish,
because in our case the intermediate propagator, connecting two boson
loops, carries the zero external momentum and is zero at any external
$\omega \neq0$, see (\ref{bare-g}).

Given ten possible decay processes (\ref{curvaEqual}), one has to
analyze $\sim 10^2$ diagrams, corresponding to Fig.\ \ref{fig:fish1}.
It is further complicated by the necessity to perform the
calculation at finite $T$ and make an analytical continuation to real
frequences $\omega \to 0$. Fortunately, there is a method which allows
the essential simplification of this task.

\subsection{Calculation by "Unitarity condition" method}

At finite $T$, the analytical continuation of the diagrams and
the essential reduction of their number under consideration are
simultaneously achieved in the "Unitarity condition" method.
\cite{Maleyev70} Since this method for $T>0$ is largely unknown to the
wider audience, we sketch the main points of it below. Readers not
interested in the details of calculation may skip to Eq.\
(\ref{r0gammas}) below.

\begin{figure}
\includegraphics[width=2cm]{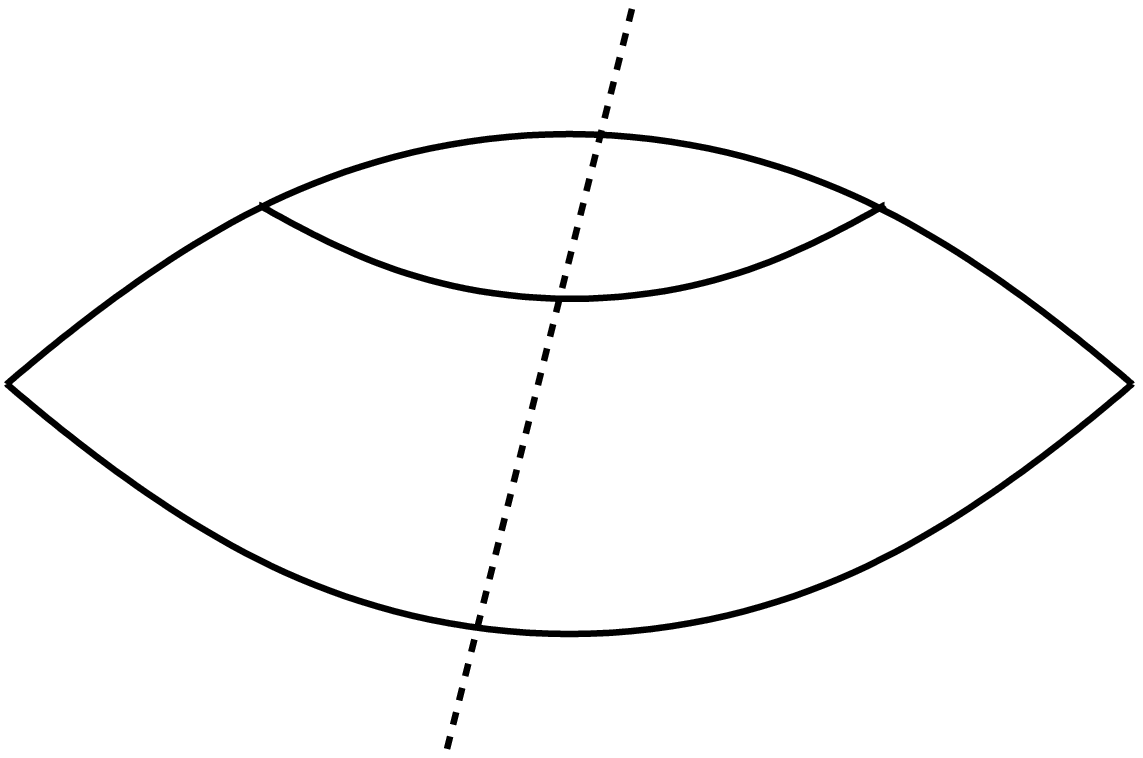}%
\hspace{.7cm}\includegraphics[width=2cm]{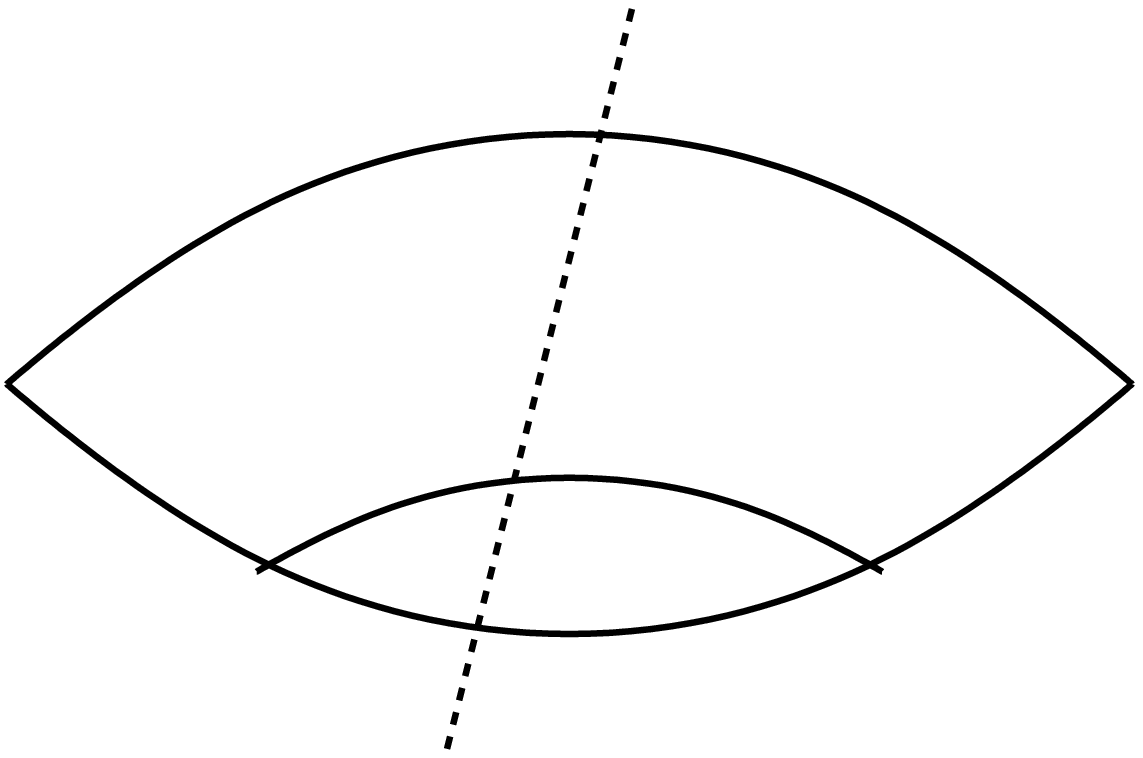}
\\
\includegraphics[width=2cm]{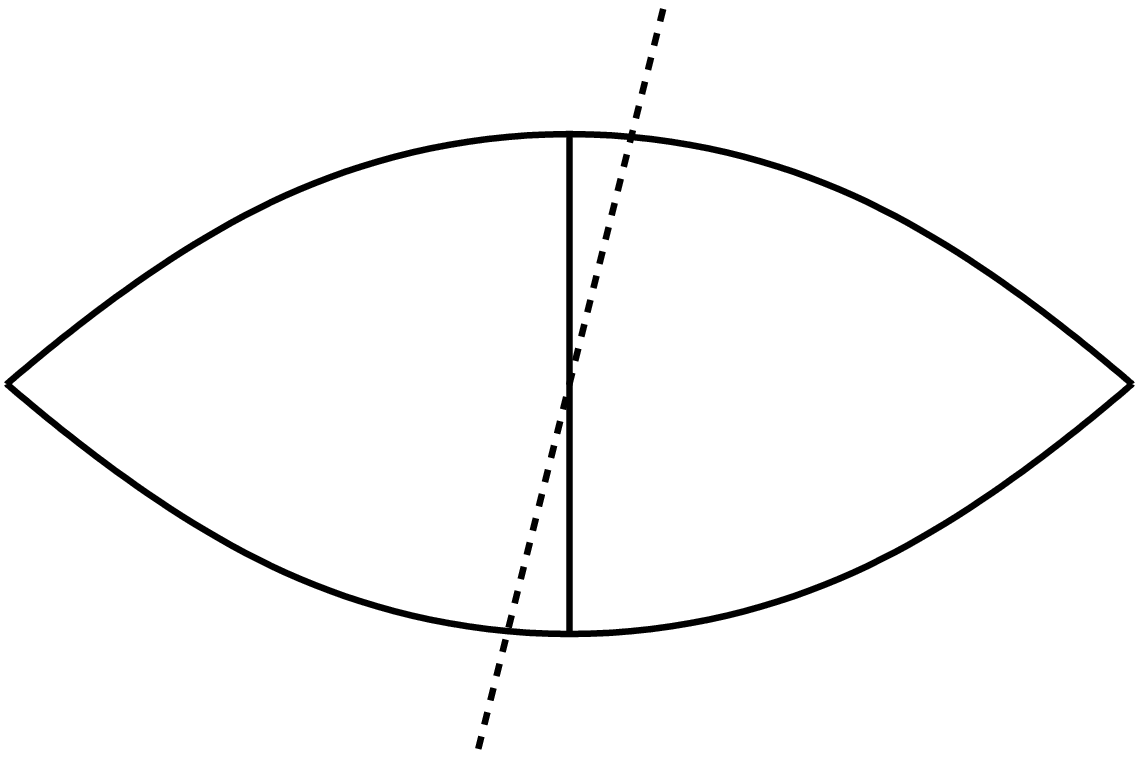}%
\hspace{.7cm}\includegraphics[width=2cm]{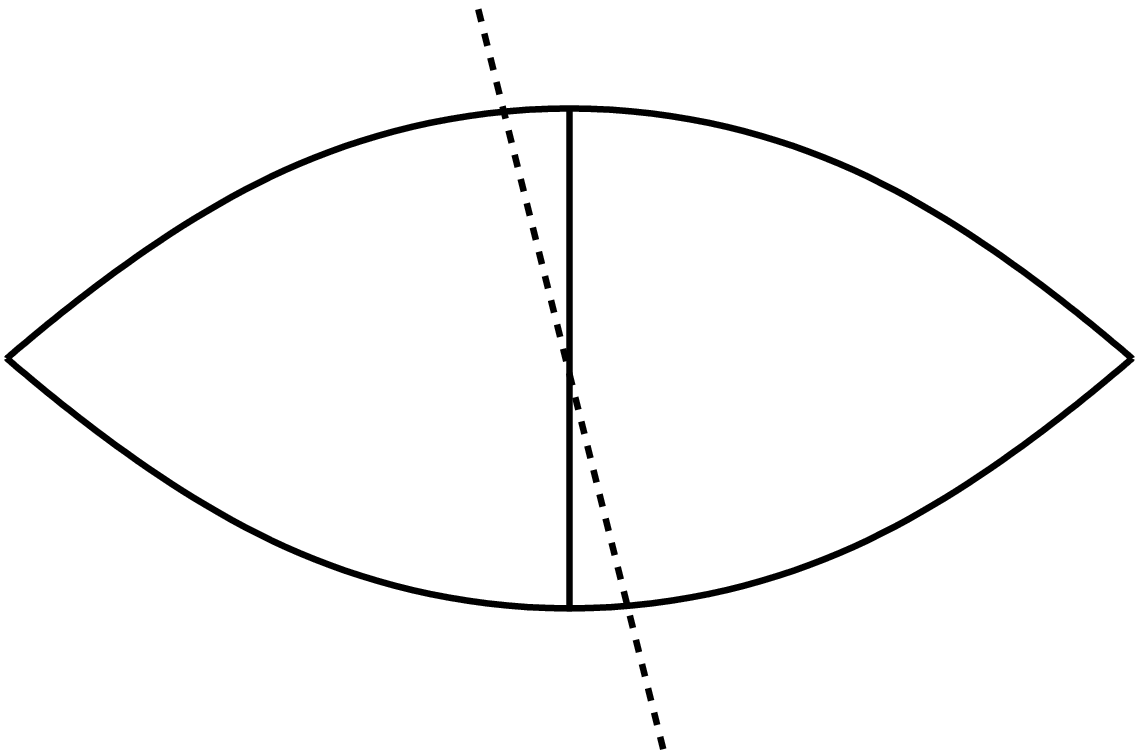}%
\caption{ Four cross sections corresponding to two above
skeleton graphs.
\label{fig:fish1sec}}
\end{figure}

\begin{figure}
\includegraphics[width=1in]{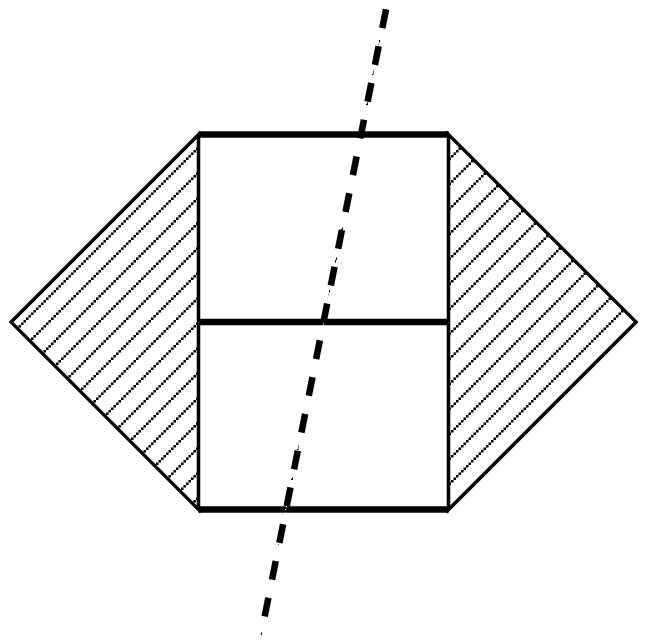}%
\caption{Leading contribution to low-T drag coefficient is given by
diagrams with three propagators in a cross section.
\label{fig:dragCS}}
\end{figure}

\begin{figure}
\includegraphics[width=0.8\columnwidth]{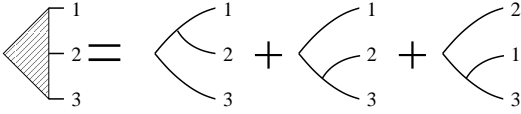}%
\caption{First-order in curvature contribution to the vertex part in
the above diagram.
\label{fig:vertex}}
\end{figure}

We define a cross section of the diagram as a line, bisecting this
diagrams in two parts, each with the external vertex $\rho_- \partial_x
\rho_+$ carrying the external frequency $\omega$ (see Fig.\
\ref{fig:fish1sec}). We should take the
product of spectral weights of the Green's functions ${\rm Im}\;
g(x_i, q_i)$ at the cross section, to divide each individual spectral
weight by $2\sinh(x_i/2T)$ (in our case of bosons), and to multiply the
whole product by $2\sinh(\omega/2T)$. The sum over all possible
cross sections, multiplied by the square of the generalized vertex
part, gives a value of the {\em imaginary part} of the
diagram. See Figures \ref{fig:dragCS} and \ref{fig:vertex},
schematically showing this method.

Using this method, all contributions can be classified by the number
and types of Green's functions appearing in the cross section. Formally,
we have the contributions with two propagators in the cross section
(2-CS) and three propagators in the cross section (3-CS). However the
cross sections with two propagators are essentially reduced to the
simple loop diagram, with a static vertex correction (except for
diagrams with $\alpha_1$ vertex, see below). Given the
structure of the external vertex, $\rho_- \partial_x \rho_+$, and
kinematic properties for linear spectrum, one can verify that the
contribution of 2-CS to $\sigma$ does not contain $\omega^{-2}$.
We may hence exclude this type of cross-section from our
analysis.

Each partial 3-CS contribution to the drag
resistivity can be represented as \cite{Maleyev70} (cf.\ Eq.\
(\ref{dragcoef}) ) :
 \begin{eqnarray}
 \delta r_0& = & \pi U_{12}^2 K_+K_-\int do \;  |\Gamma_{jkl}|^2
 S_{jkl}\label{Csection0a},
 \\
  do &=&2\pi^2 \delta\left(\sum q_i\right)
   \delta\left(\omega - \sum x_i\right)
   \prod_{i=1}^3
  \left[\frac{dq_i dx_i}{(2\pi)^2} \right]
   \label{Csection0}  ,\\
  S_{jkl} & =&
  \frac{g_{j}''(x_1,q_1) g_{k}''(x_2,q_2) g_{l}''(x_3,q_3)}
 {T\sinh\frac{x_1}{2T}\;
 \sinh\frac{x_2}{2T}\; \sinh\frac{x_3}{2T}},
 \nonumber
 \end{eqnarray}
and the spectral weight  $g_{j}''(x,q) =
\mp (q/2) \delta(qv_j \pm \omega)$ is imaginary part of the retarded
Green's function for chiral density $j = \tilde R_\pm, \tilde L_\pm$,
Eq.\ (\ref{bare-g}). A symmetrization prefactor $1/2$ should be
added to $S_{jkl}$ when values of two indices are coinciding.
The total drag resistivity is given by summation over all possible
$j,k,l = R_\pm, L_\pm$ so that

 \begin{eqnarray}
 r_0& =  &\pi U_{12}^2 K_+K_-\sum_{jkl}
 \int \frac{dq_1 dq_2 dx_1 dx_2}{(2\pi)^2} |\Gamma_{jkl}|^2
 \label{Csection}
 \\& \times&
 \frac{g_{j}''(x_1,q_1) g_{k}''(x_2,q_2)
 g_{l}''(\omega - x_1-x_2,-q_1-q_2)}
{2T \sinh\frac{x_1}{2T}\sinh\frac{x_2}{2T}
\sinh\frac{x_1+x_2 -\omega}{2T} }.
 \nonumber
 \end{eqnarray}

We can distinguish between the cross sections with i) three different
chiral densities, ii) two different chiral densities and iii) one sort
of chiral density.

I. One sort of chiral density in the cross section can appear only
on 3-CS of the type
$g_{R-}''g_{R-}'' g_{R-}'' $ and $g_{L-}''g_{L-}'' g_{L-}''$.
The kinematic restrictions (\ref{Csection0}) obviously lead to the
absence of such
contribution in (\ref{Csection}) at $\omega \neq 0$.

Having this contribution excluded, we observe that the
$\omega^{-2}$ contribution to $\sigma$ is delivered by setting
$\omega=0$ in the integrand of (\ref{Csection}); upon doing this
the calculation is simplified.

II. It can be shown that 3-CS with three different propagators
may contain only two combinations $g_{R+}''g_{L+}'' g_{L-}''$ and
$g_{R+}''g_{L+}'' g_{R-}''$. Summing up all the contributions,
and using the kinematic restrictions in $do$ and $S_{ijk}$,  Eq.\
(\ref{Csection0}), we obtain that the corresponding static vertex is
 \begin{equation}
 \Gamma_{R+L+L-} =
  \frac{q}{2\pi}\frac{4 \alpha_2  v_+^2 +
  \alpha_3  (v_+ - v_-)^2 - \alpha_4  (v_+ + v_-)^2
  }{2v_+ (v_+^2 - v_-^2)},
  \label{gamma3bosons}
 \end{equation}
here $q$ refers to $L_-$ boson ; one has $ \Gamma_{R+L+R-} =
-\Gamma_{R+L+L-}$. The contribution to the drag coefficient is
proportional to
 \begin{equation}
 \int \frac{dq\, q^3 |\Gamma_{R+L+L-}|^2 (v_+^2-v_-^2)v_+^{-3} }
  { T \sinh \frac{(v_+-v_-) q}{4T}
  \sinh \frac{(v_++v_-)q}{4T} \sinh \frac{v_- q}{2T} },
 \label{3diff-bosons}
 \end{equation}
this expression corresponds to  (\ref{dragcoef1a}). However,
substituting (\ref{alphas}) into (\ref{gamma3bosons}), one
obtains $\Gamma_{R+L+L-} = 0$ and the absence of this partial
contribution to $r_0$ (cf.\ next subsection).

III.
The remaining combinations, i.e.\  3-CS with two different
propagators,  are : $g_{R-}''g_{R-}'' g_{L-}''$,
$g_{R+}''g_{R+}'' g_{R-}''$,  $g_{R+}''g_{R+}'' g_{L-}''$ and the
combinations which are obtained from the above ones by the change
$R\leftrightarrow L$. Due to obvious symmetry, the latter change
together with $q_i \to - q_i$ leaves the expressions for $\delta r_0$
intact, so it suffices to consider only the above three CSs and
multiply the result by two.

The first cross section, $g_{R-}''g_{R-}'' g_{L-}''$, corresponds to
the non-vanishing vertex
 \begin{equation}
 \Gamma_{R-R-L-} =
 \frac{q_1}{2\pi}
 \frac{2\alpha_5 v_+}{v_+^2 - v_-^2},
 \end{equation}
where the kinematic restrictions were used again.

\begin{figure}
\includegraphics[width=0.8\columnwidth]{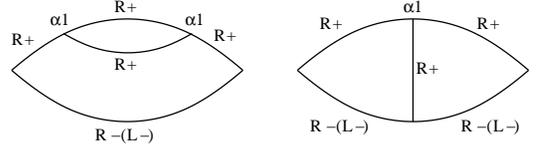}%
\caption{Singular diagrams.
\label{fig:singular}}
\end{figure}

Finally, we come to the most problematic term of the
cross sections $g_{R+}''g_{R+}'' g_{R-}''$ and  $g_{R+}''g_{R+}''
g_{L-}''$.
The corresponding vertices $\Gamma$'s in
(\ref{Csection0a}), Fig.\ \ref{fig:vertex}, explicitly read as

 \begin{eqnarray}
 \Gamma_{R+R+R-}& =  &
 \alpha_1 q_3 g_{R+}(x_1+x_2, q_1+q_2)
 \label{singGamma}
 \\ && + \alpha_2 q_3 g_{L+}(x_1+x_2, q_1+q_2)
 \nonumber \\ && - \alpha_3 q_1 g_{R-}(x_2+x_3, q_2+q_3)
  \nonumber\\ && - \alpha_5 q_1 g_{L-}(x_2+x_3, q_2+q_3)
  \nonumber , \\
  \Gamma_{R+R+L-}& =&
 \alpha_1 q_3 g_{R+}(x_1+x_2, q_1+q_2)
 \label{singGamma2}
 \\ && + \alpha_2 q_3 g_{L+}(x_1+x_2, q_1+q_2)
 \nonumber \\ && - \alpha_5 q_1 g_{R-}(x_2+x_3, q_2+q_3)
  \nonumber\\ && - \alpha_4 q_1 g_{L-}(x_2+x_3, q_2+q_3)
  \nonumber.
 \end{eqnarray}
It is seen here, that the term $\sim \alpha_1$ is
divergent on the assumption of ideal bosons, which happens because of
simultaneous conservation of energy and momentum for the linearized
spectrum. This divergence in $g_{R+}(x_1+x_2, q_1+q_2)$  would not
happen if we would assume a damping of boson excitations ;
moreover, the kinematic
restrictions in the cross section (\ref{Csection0}) lead to prefactor
$q_3 = 0 $ at $\omega =0$, so that the contribution  from $\alpha_1$
term is zero for any model damping of bosons. In our problem
however, this "damping" can only be obtained as the result of
resummation of higher order corrections in $m^{-1}$.
The divergent diagrams are depicted in Fig.\  \ref{fig:singular}.

The first diagram here, $\sim \alpha_1^2$, is the
lowest order contribution to the self-energy part, Eq.\ (\ref{Sigma}).
Evidently it corresponds to the triple pole $(x-v_+q)^{-3}$ which
resulted in the above divergence. At the same time, as discussed above,
this is the first singular diagram in a series, which should ultimately
produce the broadened propagator, Eq.\ (\ref{exacpropT}). Importantly,
this type of singular diagrams appears without the properly fermionic
interaction, as seen in (\ref{alphas}).
Therefore, in the application of the formula (\ref{Csection}), one
should omit the most singular contribution, $\sim \alpha_1^2$, in
$|\Gamma_{R+R+R-}|^2$ and use the the broadened propagator, Eq.\
(\ref{exacpropT}) instead.

The subleading singular term, $\sim \alpha_1$, is depicted by the
right diagram in Fig.\ \ref{fig:singular}. We argue here that this
diagram should vanish in our consideration, and provide the following
heuristic argument.
The linear dispersion law for bosons led to the
singularity in $\alpha_1$ term (\ref{singGamma}), but this singularity
is removed on the assumption of finite width of the bosonic pole. In
this case the Green's function $g_{R+}(x_1+x_2, q_1+q_2)$ in
(\ref{singGamma}), (\ref{singGamma2})becomes purely
imaginary quantity.  Considering the purely
real $|\Gamma_{R+R+R-}|^2 = \Gamma_{R+R+R-}^\ast\Gamma_{R+R+R-}$ one
sees that the linear-in-$\alpha_1$ terms disappear.
This argument is most obvious for the simple Lorentzian form of
$g_{R+}(x, q)$ and should be generalized for more
complicated cases, particularly, for $g_{R+}(x, q)$ given by
(\ref{exacpropT}). The analytical structure of more
complicated diagrams at finite $T$ becomes rather involved, see Ref.\
\onlinecite{Maleyev70}, and a general proof of the cancellation
of these vertex corrections should be considered elsewhere.
Here we make a conjecture that the singular terms represented by
the second diagram in Fig.\  \ref{fig:singular} vanish exactly for
the "dressed" propagators, given by (\ref{exacpropT}).

This conjecture basically amounts to the following convention :
\\  i) one should discard the curvature term $\alpha_1$ from the
Hamiltonian,
\\ ii) one should use the dressed propagators (\ref{exacpropT}) for the
modes $R_+, L_+$, with the "effective mass" $1/m^\ast =
\alpha_1/(2\pi^2)$, and
\\  iii) the other curvature terms $\alpha_2, \ldots, \alpha_5$, not
leading to singularities, can be treated in the regular PT.
\\
Proceeding this way, we conclude that at higher temperatures, $T \agt
T_1$ we can use Eq.\ (\ref{dragcoef1a}) with the change ${\rm Im}\;
\chi_1 {\rm Im}\;
\chi_2 \to {\rm Im}\; \chi_+ {\rm Im}\; \chi_-$. The finite width of the
dressed
propagators $g_{R+} , g_{L+}$ leads then to the regime $r_0 \sim T^2$,
similar to Eq.\ (\ref{simpleT2}).

The interaction-induced corrections \cite{PKKG06} to the shape of
propagator (\ref{exacpropT}) are the next-order effect in curvature,
and do not influence our calculation of $r_0$ below. Indeed, at lowest
temperatures,  $T\ll T_1$, the peaks in spectral weights ${\rm Im}\;
\chi_i$ in (\ref{dragcoef1a}) do not overlap, and it suffices to
consider the "tail" of one propagator in (\ref{dragcoef1a}) and the
core of another one, approximated by $\delta-$function. At higher
temperatures, $T\agt
T_1$, propagators in (\ref{dragcoef1a}) significantly overlap; this is
non-perturbative regime for bosonization, which cannot provide a
prefactor in the estimate $r_0 \sim T^2$. However, one may argue that
all interaction-induced details \cite{PKKG06} in the shape of the core
of propagators are smeared by higher temperatures and one can safely
use the free-fermion estimate (\ref{exacpropT}) here.

At lower temperatures, $T \alt T_1$, we obtain the finite contribution
$r_0 \sim T^5$, due to vertices  $\alpha_2, \ldots, \alpha_5$, absent
in the non-interacting situation. A rather long and straightforward
calculation gives a prefactor in this $T^5$ temperature dependence :

      \begin{eqnarray}
      r_0 &=& U_{12}^2\gamma T^5 , \quad
      \gamma = \gamma_1 + \gamma_2 + \gamma_3 ,
      \label{r0gammas}
       \\
       \gamma_1&=&\frac{1}{4\pi^2}
       \frac{\left[\alpha_3(v_+ + v_-) -
       \alpha_5(v_+ - v_-) \right]^2}
       {30v_+^5v_-(v_+ - v_-)(v_+^2 - v_-^2)^2},
       \nonumber \\
        \gamma_2&=&\frac{1}{4\pi^2}
        \frac{ \left[\alpha_5(v_+ + v_-) -
       \alpha_4(v_+ - v_-) \right]^2}
       {30v_+^5v_-(v_+ + v_-)(v_+^2 - v_-^2)^2},
       \nonumber \\
        \gamma_3&=&\frac{1}{4\pi^2}
        \frac{\alpha_5^2 v_+^2}
       {15v_-^7(v_+^2 - v_-^2)^2},
       \nonumber
      \end{eqnarray}
with $\alpha_j$ given by (\ref{alphas}). It is seen that at $v_+ - v_-
\to 0$ the leading term in $\gamma$ is given by $\alpha_3^2$ term and
is $\propto (v_+ - v_-)^{-3} $, which is different from
the above regime $ \sim (v_+ - v_-)^{-2}$ non-equal wires, Eq.\
(\ref{dragcoef1b}). To resolve the seeming discrepancy, we perform the
analysis of the nearly-equal wires below.

\subsection{Nearly-identical wires}

We consider now the case when the plasmon velocities $v_{1,2}$ in
individual wires are nearly coinciding and their difference is of order
of $U_{12}$. The main idea of this subsection is that the
coefficients in the canonical transformation diagonalizing the quadratic
Luttinger Hamiltonian are sensitive to the ratio $U_{12}/(v_1 - v_2)$.
These coefficients, in turn, define the values $\alpha_i$ for the
vertices (\ref{curvaEqual}). We can calculate these values and use the
previous expressions (\ref{3diff-bosons}), (\ref{r0gammas}) to obtain
the drag in this more general case.

Using dimensionless difference of plasmon velocities (\ref{defDelta}),
it is convenient to introduce the parameter $\lambda$ as follows :
        \begin{equation}
        u = \Delta \sin \lambda , \quad
        v_1 - v_2 \simeq v \Delta \cos \lambda,
        \end{equation}
where the first equality is the properly definition of $\lambda$.

The general Bogoliubov transformation, diagonalizing the quadratic
part of the Hamiltonian (\ref{ham}), explicitly reads as
        \begin{eqnarray}
        \begin{pmatrix}
        \theta_1 \\  \theta_2 \\ \phi_1 \\ \phi_2
        \end{pmatrix} &=&
         S_1  . S_2 .  S_3 .
          \begin{pmatrix}
        \theta_+ \\  \theta_- \\ \phi_+ \\ \phi_-
        \end{pmatrix},
        \label{uv-Bogol}
        \end{eqnarray}     with          \begin{eqnarray}
         S_1 &=& diag\left[ v_{J1}^{-1/2} ,  v_{J2}^{-1/2},
         v_{J1}^{1/2} ,  v_{J2}^{1/2}
         \right]      \nonumber   ,\\
         S_3 &=& diag\left[ v_{+}^{1/2} ,  v_{-}^{1/2},
         v_{+}^{-1/2} ,  v_{-}^{-1/2}
         \right]      \nonumber   ,\\
         S_2 &=&
        \begin{pmatrix}
        \cos (\lambda/2) ,& - \sin (\lambda/2) ,& 0 ,& 0 \\
        \sin (\lambda/2) ,& \cos (\lambda/2) ,&  0 ,& 0 \\
         0 ,& 0 ,&  \cos (\lambda/2) ,& - \sin (\lambda/2)  \\
         0 ,& 0 ,& \sin (\lambda/2) ,& \cos (\lambda/2)
        \end{pmatrix} \nonumber,
        \end{eqnarray}
where $v_{+}$ and $v_{-}$ are given by (\ref{velocities}) and correspond
to the new modes $\phi_{+}$ and $\phi_{-}$, respectively. Using this
transformation, we can determine the values of the coefficients
$\alpha_i$ in (\ref{curvaEqual}).

Note that the set of the vertices (\ref{curvaEqual}) was obtained for
the case $\lambda  = \pi /2$. For other values of $\lambda$, we
have also another set of the vertices which is obtained from
(\ref{curvaEqual}) by a change $R_+ \leftrightarrow  R_-$ and $L_+
\leftrightarrow  L_-$, explicitly
    \begin{eqnarray}
     H_{cur}' &=&
     \frac{\alpha'_{1}}{3} \left(
     \tilde R_-^3 + \tilde L_-^3\right) +
      \frac{\alpha'_{2}}2
     \left( \tilde R_-^2 \tilde L_- +
     \tilde R_- \tilde L_-^2 \right)
     \nonumber \\  && +
      \frac{\alpha'_3}2
      (\tilde R_-\tilde R_+^2 + \tilde L_-\tilde L_+^2)
     +  \frac{\alpha'_4}2
     (\tilde R_-\tilde L_+^2 + \tilde L_-\tilde R_+^2)
     \nonumber \\  &&
     + \alpha'_5 ( \tilde R_- + \tilde L_- ) \tilde R_+ \tilde L_+
     \label{curvaClose}.
    \end{eqnarray}

For our purposes here, it is sufficient to let $v_{J1} = v_{J2}$ and
$s_{1} \simeq s_{2}$ in the calculation of $\alpha_j, \alpha'_j$ . Thus
we neglect the difference in plasmon velocities in (\ref{uv-Bogol}). At
the same time, we allow for the intrawire interaction, $K = v_{J1} /
s_{1} \neq 1$.

We obtain the expressions for vertices in (\ref{curvaEqual}) as
   \begin{eqnarray}
    \alpha_1 &\simeq& \frac{\pi^2}{m} \frac{3+K^2}{2\sqrt{K}}
    \left( \cos^3 \frac\lambda2 +  \sin^3 \frac\lambda2 \right)
    \nonumber ,\\
    \alpha_2 &\simeq&\frac{\pi^2}{m} \frac{K^2 - 1}{\sqrt{K}}
    \left(\cos^3 \frac\lambda2 +  \sin^3 \frac\lambda2 \right)
    \nonumber,\\
    \alpha_3 &\simeq& \frac{\pi^2}{m}\frac{K^2 + 3}{2\sqrt{K}}
    \left(\cos \frac\lambda2 +  \sin \frac\lambda2\right)
    \sin  \lambda
    \label{generalAlphas},
    \\
    \alpha_4 &\simeq&\frac{\pi^2}{m} \frac{K^2 -1}{2\sqrt{K}}
    \left(\cos \frac\lambda2 +  \sin\frac\lambda2\right)
    \sin  \lambda
    \nonumber ,\\
    \alpha_5 &\simeq&  \alpha_4
    \nonumber,
    \end{eqnarray}
and the set $\alpha'_j$ is obtained from (\ref{generalAlphas}) by
changing $\alpha_j \to \alpha'_j$ and $\lambda \to -\lambda$.

One can verify that despite more complicated structure
(\ref{uv-Bogol}), the drag current vertex is still given by the last
expression in (\ref{ext-vertex}).
Further, the existence of the second set of vertices,
$\alpha'_j$, leads to the same set of diagrams as before, with
appropriate change $R_+ \leftrightarrow R_-$ etc. Importantly,
there are no diagrams of the type Fig.\ \ref{fig:fish1}, containing
both $\alpha'_j$ and $\alpha_j$.  It means that one can use the previous
expressions (\ref{r0gammas}), corroborated by the previously vanishing
contribution (\ref{3diff-bosons}) and to add the contribution from the
$\alpha'_j$ set, obtained by $\lambda \to -\lambda$.

Combining all terms, we find
      \begin{eqnarray}
      r_0 &\simeq& \frac{\pi^3 K U_{12}^2 T^5}{120m^2 v^9}
      \left[
      \frac{(1-K^2)^2\cos^2 \lambda}{\Delta^2}
      +  \frac{(3+K^2)^2\sin^2 \lambda}{4\Delta^3}
      \right] \nonumber \\
       &\sim &\frac{ U_{12}^2 T^5}{E_F^2 v_F^5}
      \left( \frac{\cos^2 \lambda}{\Delta^2} O(1)  +
      \frac{\sin^2 \lambda}{\Delta^3} O(1)
      \right).
      \label{finalT5}
      \end{eqnarray}
This is the central result of this section. Let us discuss it in more
detail.

First of all, we see that the term $\sim \Delta^{-2}$ in
(\ref{finalT5}) is proportional to intrawire interaction strength,
$(1-K)^2$, whereas the term $\sim \Delta^{-3}$ is present even
for initially non-interacting wires. This is not surprizing, because  in
the latter case the factor $\sin^2 \lambda \sim U_{12}^2/ (v_1-v_2)^2$
means that $U_{12}$ serves as the Luttinger-type interaction $(K-1)$,
this term should be obtained for non-equal wires in the order $U_{12}^4$
of conventional fermionic formalism.

For identical wires, we have an interesting crossover from $U_{12}^2
T^2$ regime at high temperatures to $T^5/|U_{12}|$ behavior at lower
temperatures. Recalling our definitions of the drag resistivity $R_{12}
= \pi r_0/(k_{F1} k_{F2})$ and of the small scale $T_1 = E_F \Delta \sim
U_{12} k_F$, we obtain
      \begin{eqnarray}
      R_{12} &\sim&
      \frac{T^5}{v_F E_F^3 T_1 }
      ,\quad T<T_1 \sim U_{12} k_F ,
      \label{T5T2} \\
       &\sim&
        \frac{ T_1^2 T^2}{v_F E_F^3 }
      ,\quad T>T_1 ,    \nonumber
      \end{eqnarray}
which shows a smooth crossover at $T\simeq T_1$.  The parametrically
large prefactor, $1/U_{12}$, is similar to above
Eq. (\ref{cond.ident.wires}). This prefactor i) shows non-perturbative
character of the obtained result and ii) does not mean actual
enhancement of the drag effect, due to its overall small amplitude
$\sim T^5$.

For the wires with different plasmon velocities we have $T_1
\simeq (v_1-v_2) k_F$, and the above Eq.\ (\ref{finalT5}) shows the
following behavior

      \begin{eqnarray}
      R_{12} &\sim&
      \frac{T^5}{v_F E_F^2 T_1^2 }\frac{U_{12}^2 (1-K)^2}{v_F^2}
      ,\quad T<T_1  ,
      \label{T5T2dif} \\
       &\sim&
        \frac{U_{12}^2 T^2}{v_F^3 E_F }
      ,\quad T>T_1 .    \nonumber
      \end{eqnarray}
and two regimes do not match at $T = T_1$. As we saw above, $T^2$
regime changes to exponential decrease at $T<  T_1$,
which means that there is a logarithmically narrow crossover
region of temperatures, where  $R_{12}$ interpolates between $T^5$ and
$T^2$ in (\ref{T5T2dif}).

\section{Summary and conclusions}
\label{sec:discussion}

In this paper we discuss the effects of curvature for the Luttinger
liquid. The curvature is irrelevant perturbation, in the sense that it
does not change the low-energy of the system and usually can be
simply discarded. Bosonization technique regards the curvature as
interaction and, at the first glance, a regular perturbation theory is
possible. However, a certain care must be exercised with such
PT, in spite of evident correspondence between fermionic
and bosonization results.  We show, that the PT in curvature in
bosonization produces both singular and regular contributions to
different quantities. The singular contributions arise already for the
free fermions and correspond to the non-Lorentzian character of
fermionic density propagator. We show that these singular contributions
stem from one vertex of boson interaction, and we propose to exclude
this vertex from PT analysis, while simultaneously using the
free-fermion expression as a "dressed" form of bosonic propagators.
For the Coulomb drag problem,
the dressed form of propagators becomes important at high temperatures.

All other bosonic vertices appear due to
fermionic interaction and yield regular contributions in bosonization
PT. They are responsible for the low-$T$ regime in the Coulomb drag
problem.

Comparing bosonization to fermionic approach at the example of Coulomb
drag problem, we observe a complementary character of these methods.
High-temperature $T^2$ regime is obtained most easily with fermions, but
this is essentially non-perturbative regime for bosonization.
Conversely, the low-temperature $T^5$ drag regime for different wires
is obtained relatively easily with bosons, whereas the
calculation by fermionic approach is cumbersome. We demonstrated that
the bosonization is particularly useful in case of Coulomb drag between
nearly-equal wires, when it simultaneously resums several RPA
channels for fermionic diagrams.

We did not include spin into our consideration. It also
requires a separate analysis, because the cubic terms in the fermionic
densities violate the spin-charge separation realized at the level of
quadratic action.
It is evident that spinful electrons in one wire are
equivalent to spinless fermions in two wires. In our formalism, the
Hamiltonian is given by (\ref{hamsymm}) with identification $\phi_+$
($\phi_-$) for charge (spin) density.
The above case of nearly identical wires corresponds to
Zeeman-splitting of the Fermi velocities in the magnetic field by
value $\sim B/k_F$.

Finally, we note that in one spatial dimension, the alternative source
for the "ideal" drag was proposed by Nazarov and Averin. \cite{NazAve98}
They considered  the "backscattering" between the wires, i.e. $2k_F$
Fourier component of the interwire interaction $U_{12}(2k_F)
=U_{bs}$, not considered here. The effect of renormalization leads
to the increase $\bar U_{bs} = U_{bs} (E_F/T)^{2-2K}$ with
lowering $T$, and the drag resistivity $R_{12} \sim \bar
U_{bs}^2$ diverges in the limit $T\to 0$.
However, as was noted in \cite{PMGA,Klesse00}, for large interwire
distances $D$, the backscattering amplitude is small
$U_{12}^{bs} \sim e^{-2k_F D}$ and the effect of renormalization can
compensate this smallness only at exponentially small temperatures.
The backscattering drag mechanism is also sensitive to the difference
of fermionic densities in two wires. \cite{NazAve98,Klesse00}
Particularly, a straightforward calculation by Fuchs et al.\
\cite{Klesse00} showed the exponential suppression of $R_{12}$ due to
backscattering at $T < v_F (k_{F1}-k_{F2}) = T_1^\ast$.
In view of an obvious similarity between $T_1^\ast$ and our $T_1$, it is
tempting to regard the regime $R_{12} \sim T^5$ as the only one,
surviving at low temperatures. The inclusion of the backscattering
drag mechanism into our discussion is clearly beyond the scope of this
study.

\begin{acknowledgments}
I thank L.I.\ Glazman, M. Pustilnik, A. Luther, A. Kamenev, D.G.\
Polyakov, S. Teber, A.G.\ Yashenkin for numerous useful discussions.
I am very grateful to I.V.\ Gornyi and A.D.\ Mirlin for intensive
discussions at the final stage of this work and valuable comments on
the draft.
\end{acknowledgments}

\appendix*

\section{optical conductivity}
We calculate optical trans-conductivity according to Kubo formula,
(\ref{Kubo}) and have ${\rm Re}\; \hat\sigma_{jl} = \omega^{-1} {\rm
Im}\; Q_{jl}(\omega)$ where $Q_{jl}(\omega)$ is a retarded current
response function.  Combining different contributions to $\hat
Q(\omega)$, depicted in Fig. \ref{fig:dragloop} and making analytical
continuation to real frequencies at finite $T$, we obtain
    \begin{eqnarray}
    Q_{jl} &= &
    \int dq dx_1 dx_2 F_{jl}
    \frac{N(x_1) - N(x_2)}{\omega- x_1 + x_2 +i 0}
    W(x_1) W(x_2)
    \nonumber , \\
     F_{12}&= & F_{21} =
     \frac 12 q^6 u^2v_1^2 v_2^2 (x_1+x_2)^2 ,\\
     F_{11}&= &
     q^2 (x_1^2 -v_2^2 q^2)(x_2^2 -v_2^2 q^2)(x_1x_2 +v_1^2 q^2)
       \nonumber \\ &&
      + \frac 12  q^6 u^2v_1^2 v_2^2 (x_1^2+x_2^2 - 2v_2^2 q^2),
      \\
          W(x) &=& \frac{1}{2x(\varepsilon_1^2 - \varepsilon_2^2)}
    \left( \delta(x+\varepsilon_1) +\delta(x-\varepsilon_1)
    \right.  \nonumber \\ && \left.
    -    \delta(x+\varepsilon_2) -\delta(x-\varepsilon_2)
    \right),
     \label{denomG}
    \end{eqnarray}
and $F_{22}$ is obtained from $F_{11}$ by interchanging $v_1$ and $v_2$
; here $N(x)$ is Planck function.

We calculate ${\rm Im}\; Q_{12}$ at $\omega\neq 0$. For non-zero
frequencies the coinciding arguments $x_1 = x_2 = \pm
\varepsilon_{1,2}$ do not
contribute and the rest of the integration is simple. After collecting
all the terms coming from (\ref{denomG}) we obtain the real part of the
conductivity ${\rm Re}\; \hat\sigma (\omega)$ in the form
(\ref{sigmaGen}), with $\sigma_d$ given by (\ref{opt.condu.finiteT}).



\begin{thebibliography}{99}

\bibitem{PMGA} M. Pustilnik, E.G. Mishchenko, L.I. Glazman, and  A.V.
Andreev, Phys. Rev. Lett. 91, 126805 (2003).

\bibitem{AbaWieg05} A. G. Abanov and P. B. Wiegmann,
Phys. Rev. Lett. {\bf 95}, 076402 (2005).

\bibitem{G7} R. G. Pereira, J. Sirker, J.-S. Caux, R. Hagemans, J.
M. Maillet, S. R. White, and I. Affleck,  Phys. Rev. Lett. 96, 257202
(2006)

\bibitem{Teber06} S. Teber, Eur. Phys. J. B 52, 233
(2006); cond-mat/0609754.

\bibitem{PKKG06} M. Pustilnik, M. Khodas, A. Kamenev, and L. I. Glazman,
 Phys. Rev. Lett. 96, 196405 (2006).

\bibitem{Kopietz05} P. Pirooznia, P. Kopietz, cond-mat/0512494

\bibitem{Gutman06}
D.B. Gutman,  cond-mat/0608448 ; cond-mat/0612234

\bibitem{reviews}
A.G.\  Rojo, J. Phys.: Condens. Matter {\bf 11} R31 (1999);
P. Debray, V. N. Zverev, V.L. Gurevich, R. Klesse and R. S. Newrock,
Semicond. Sci. Technol. {\bf 17},  R21 (2002).

\bibitem{drag90s}
A.-P. Jauho and H. Smith, Phys. Rev. B 47, 4420 (1993);
L. Zheng and A.H. MacDonald, Phys. Rev. B 48, 8203 (1993);
K. Flensberg, B. Yu-Kuang Hu, A.-P. Jauho and  J.M. Kinaret,
Phys. Rev. B 52, 14761 (1995);
A. Kamenev and Y. Oreg, Phys. Rev. B 52, 7516 (1995).

\bibitem{GoMivOp04} I. V Gornyi, A. D. Mirlin, and F. von Oppen, Phys.
Rev. B 70, 245302 (2004).

\bibitem{NazAve98} Yu.V.\ Nazarov and D.V.\  Averin,
Phys.Rev.Lett.\ {\bf 81}, 653 (1998).

\bibitem{GoNeTs} A.O. Gogolin, A.A.
Nersesyan, and A.M. Tsvelik, Bosonization and Strongly Correlated
Systems (Cambridge University Press, 1998).

\bibitem{Haldane81} F.D.M. Haldane, J.Phys. C, {\bf 14}, 2585 (1981)
\bibitem{Schick68}  M. Schick, Phys. Rev. 166, 404 (1968).
\bibitem{Tomonaga} S. Tomonaga, Prog. Theor. Phys., {\bf 5}, 544 (1950).

\bibitem{DzyaLar73} I.E. Dzyaloshinskii and A.I. Larkin,
JETP, {\bf 38}, 202 (1974).

\bibitem{Samo98}
K.V. Samokhin, J. Phys. Condens. Matter {\bf 10}, L533 (1998).

\bibitem{LuttKohn55} J. M. Luttinger and W. Kohn, Phys. Rev. 97, 869
(1955)
\bibitem{GiaMil92} T. Giamarchi and A. J. Millis, Phys. Rev. B 46, 9325
(1992)

\bibitem{AriZey05} D.N. Aristov and R. Zeyher,
Phys. Rev. B {\bf 72}, 115118 (2005).

\bibitem{GoeWoe72} W. G\"otze and P. W\"olfle Phys. Rev. B 6, 1226
(1972)

\bibitem{AmiVig00} I. D'Amico and G. Vignale, Phys. Rev. B 62, 4853
(2000)

\bibitem{Maleyev70} S.V. Maleyev, Theor. Math. Phys. {\bf 4}, 694 (1970)

\bibitem{Klesse00} R. Klesse and A. Stern,
Phys.Rev.\ B {\bf 62}, 16 912 (2000);
T. Fuchs, R. Klesse, and A. Stern,
{\em ibid.} {\bf71}, 045321 (2005).

\end{thebibliography}
\end{document}